\documentclass[apj]{emulateapj}

\newcommand{\co}{\mbox{\rm CO}}

\newcommand{\hi}{\mbox{\rm \ion{H}{1}}}

\newcommand{\htwo}{\mbox{\rm H$_2$}}
\newcommand{\jone}{\mbox{($1\rightarrow0$)}}
\newcommand{\jtwo}{\mbox{($2\rightarrow1$)}}

\newcommand{\sdunits}{\hbox{M$_\odot$~pc$^{-2}$}}

\newcommand{\kmpers}{\mbox{km~s$^{-1}$}}

\newcommand{\xcounits}{\mbox{cm$^{-2}$ (K km s$^{-1}$)$^{-1}$}}

\newcommand{\xco}{\mbox{\rm X$_{\rm CO}$}}

\shorttitle{Molecular Gas in IC~10}
\shortauthors{Leroy et al.}
\slugcomment{Submitted for publication in \emph{The Astrophysical
Journal}}

\begin{document}
\title{Molecular Gas in the Low Metallicity, Star Forming Dwarf IC 10}

\author{A. Leroy, A. Bolatto, F. Walter, L. Blitz}

\affil{Department of Astronomy, 601 Campbell Hall, University of
       California at Berkeley, CA 94720} 

\affil{Max-Planck-Institute for Astronomy, Königstuhl 17, D-69117
       Heidelberg, Germany} 

\begin{abstract}
We present a complete survey of \co\ \jone\ emission in the Local
Group dwarf irregular IC~10. The survey, conducted with the BIMA
interferometer, covers the stellar disk and a large fraction of the
extended \hi\ envelope with the sensitivity and resolution necessary
to detect individual giant molecular clouds (GMCs) at the distance of
IC~10 (950 kpc). We find 16 clouds with a total CO luminosity of $1
\times 10^6$ K km s$^{-1}$ pc$^2$, equivalent to $4 \times 10^6$
M$_{\odot}$ of molecular gas using the Galactic CO-to-H$_2$ conversion
factor. Observations with the ARO 12m find that BIMA may resolve out
as much as 50\% of the CO emission, and we estimate the total CO
luminosity as $\sim 2.2 \times 10^6$ K km s$^{-1}$ pc$^2$. We measure
the properties of 14 GMCs from high resolution OVRO data. These clouds
are very similar to Galactic GMCs in their sizes, line widths,
luminosities, and CO-to-\htwo\ conversion factors despite the low
metallicity of IC~10 ($Z \approx 1/5 Z_{\odot}$). Comparing the BIMA
survey to the atomic gas and stellar content of IC~10 we find that
most of the \co\ emission is coincident with high surface density
\hi. IC~10 displays a much higher star formation rate per unit
molecular (\htwo) or total ($\hi+\htwo$) gas than most galaxies. This
could be a real difference or may be an evolutionary effect --- the
star formation rate may have been higher in the recent past.
\end{abstract}

% INTRODUCTION
\section{Introduction}

Star forming dwarf galaxies tend to have low metallicities, intense
interstellar radiation fields, and shallow potential wells. For these
reasons, dwarfs are often used as astrophysical laboratories in which
to study the effect of extreme conditions on the interstellar medium
(ISM) and star formation. The similarity between local dwarf galaxies
and the first star forming systems --- which were also low mass and
chemically primitive --- further motivates such
studies. Unfortunately, the smallest actively star forming systems are
not very luminous and are therefore difficult to observe at any
significant distance. Systems the size of the Large Magellanic Cloud
(LMC) may be studied in some detail out to 10 Mpc or more, but
detailed observations of smaller systems are possible only in the
Local Group. In practice this limits such studies to a handful of
systems, the Local Group irregular galaxy IC~10, the Small Magellanic
Cloud (SMC), NGC~6822, and perhaps a few other Local Group dwarfs
\citep[][]{MATEO98}.

In this paper we present a new study of the molecular gas component of
IC~10. We have conducted a complete survey of $^{12}$CO \jone\
emission from IC~10 using the BIMA interferometer. This survey covers
the optical disk and much of the extended \hi\ structure surrounding
IC~10 with the resolution and sensitivity necessary to detect
individual giant molecular clouds (GMCs). We also present new single
dish observations towards most of the \co\ emission from IC~10 using
the ARO 12m and we measure macroscopic properties of $14$ GMCs using
the high-resolution OVRO observations by \citet[][]{WALTER03}. We
combine these data with literature observations of IC~10 at several
wavelengths to address three questions: 1) How do the GMCs in IC~10
compare to the GMCs found in Local Group spirals?  2) Can we predict
the molecular gas content from the hydrostatic pressure? 3) Do stars
form out of molecular gas at the same rate in IC~10 as in spiral
galaxies?

There are several reasons to expect the relationship between atomic
gas, molecular gas, and star formation in IC~10 might be different
from that in spiral galaxies. Dust plays a crucial role in setting the
abundance of molecular gas by shielding molecular gas from
dissociating radiation and serving as the site of molecular hydrogen
formation. With its low metallicity \citep[$Z_{IC~10} \approx
1/5~Z_{\odot}$,][]{GARNETT90}, IC~10 might be expected to have to have
a low dust-to-gas ratio. The interstellar radiation field (ISRF) in
IC~10 should also be more intense than in most spiral galaxies due to
the high star formation rate, low dust abundance (lower extinction),
and low metallicity (less line blanketing). A more intense ISRF will
dissociate molecular hydrogen faster and perhaps inhibit the formation
of \htwo\ by heating the grain surfaces. It may also affect different
molecular species differently as, for example, H$_2$ self-shields
while CO is dissociated. Other factors, for instance the lack of shear
(perhaps diminishing rotational support against GMC collapse), the
absence of spiral density waves (the sites of GMC formation in
spirals), and the simple underabundance of carbon and oxygen (and
thus, presumably, of CO to trace and cool molecular gas), further
distinguish IC~10 from the Galactic environment.

This paper is organized in the following way. In \S2 we summarize the
macroscopic properties of IC~10 and present maps of several components
of the galaxy. We emphasize the similarity of IC~10 to the SMC in
gross properties and discuss IC~10's vigorous present day star
formation. In \S3 we present our BIMA and ARO 12m observations,
discuss how we identified signal in the survey, describe our algorithm
for measuring GMC properties, and note several other datasets used in
this paper. In \S4 we present the results of the BIMA survey and our
GMC property measurements. In \S5 we present our analysis of the
molecular content in IC~10. We emphasize the surprising similarity
between IC~10 clouds and GMCs in large spiral galaxies. We examine
quantitative relationships between atomic gas, molecular gas, and star
formation. We consider the hypothesis that the hydrostatic pressure
predicts the molecular to atomic gas ratio and look at the efficiency
with which molecular gas forms stars in IC~10. In \S6, we summarize
our findings and suggest some future avenues of investigation in
IC~10.

% DESCRIPTION OF THE GALAXY
\section{Description of IC~10}

Along with the SMC, IC~10 is the best example of a low-mass,
metal-poor, actively star-forming galaxy in the Local Group. Table
\ref{IC10PROPS} summarizes a number of properties of IC~10. IC~10 has
stellar, atomic gas, and dynamical masses of $\sim 4,~2,~\mbox{and}~15
\times 10^{8}$ M$_{\odot}$, respectively
\citep[][]{2MASSLGA,HUCHTMEIER88,MATEO98}; by comparison, the SMC has
stellar, atomic gas, and dynamical masses of $\sim 4,~4,~\mbox{and}~24
\times 10^{8}$ M$_{\odot}$ \citep[][]{STANIMIROVIC04}. IC~10 has a
metallicity of 12 + log O/H $\approx 8.25$
\citep[][]{LEQUEUX79,GARNETT90}, intermediate between the SMC and the
LMC \citep[$8.0$ and $8.4$, respectively,][]{DUFOUR84}. Like the SMC,
IC~10 has irregular morphology, ongoing high mass star formation, and
an extended \hi\ envelope. IC~10 is probably associated with M~31 but
the separation between the two galaxies is fairly large. At our
adopted distance ($950$ kpc), IC~10 is separated from M~31 galaxy by
$\gtrsim 250$ kpc. Another close cousin to IC~10 exists just beyond
the Local Group --- the post starburst dwarf NGC~1569 at a distance of
$\sim 2$ Mpc. That galaxy is also relatively isolated, with atomic gas
and total masses of $\sim 1$ and $\sim 3 \times 10^8$
\citep[][]{ISRAEL88}, respectively, and a metallicity of $\sim
1/4~Z_{\odot}$ \citep[][]{CALZETTI1569}.

IC~10 has a higher star formation rate (SFR) than the SMC, though the
exact SFR is somewhat uncertain. The H$\alpha$ flux implies a SFR of
$\sim 0.2$ M$_{\odot}$ yr$^{-1}$ with a standard correction for
internal extinction, but this value may be as high $0.6$ M$_{\odot}$
yr$^{-1}$ if the larger extinction estimates in the literature are
correct \citep[][]{GILDEPAZ03,YANG93,BORISSOVA00}. By comparison,
\citet[][]{WILKE04} estimates an SFR of $\sim 0.05$ M$_{\odot}$
yr$^{-1}$ in the SMC (with the FIR and H$\alpha$ in agreement after
corrections for extinction and absorption). \citet[][]{MASSEY02},
\citet[][]{CROWTHER03}, and others have noted IC~10's prodigious
content of Wolf Rayet (WR) stars. More than $30$ have been
spectroscopically confirmed \citep[][]{CROWTHER03} and
\citet[][]{MASSEY02} estimate from photometry that the total number
may be $\sim 100$. By contrast, the SMC contains only $12$ WR stars
\citep[][]{MASSEY03}. The number of WR stars in a galaxy should
provide a distance independent estimate of the high mass stellar
content (and thus formation rate). Thus H$\alpha$ and WR star counts
suggest IC~10 to have a star formation rate $3$ -- $4$ times that of
the SMC. Measurements of internal extinction
\citep[][]{YANG93,BORISSOVA00} and the higher number of WR stars
estimated by \citet[][]{MASSEY02} suggest the SFR may easily be as
high as $\sim 0.5$ M$_{\odot}$ yr$^{-1}$. We note that the FIR and
radio continuum yield SFR estimates considerably lower \citep[$\sim
0.05$ M$_{\odot}$ yr$^{-1}$][]{THRONSON90,WHITE92,BELL03} perhaps as a
result of a higher UV escape fraction or lower dust abundance.

The unusually high content of WR stars has led to claims that IC~10 is
the nearest starburst galaxy --- the surface density of WR stars is
the highest of any local group galaxy and the average across the
entire galaxy is comparable to the most actively star forming regions
in M~33 \citep[][]{MASSEY95}. The disrupted morphology of the \hi\
distribution also suggests a violent recent history. The \hi\
distribution across the disk of the galaxy is characterized by seven
large holes, possibly carved out by winds or supernovae
\citep[][]{WILCOTS98}. The central regions of these holes are free of
\hi\ emission down the to a sensitivity limit of $1 \times 10^{19}$
cm$^{-2}$ (3$\sigma$), equivalent to about $0.1$ M$_{\odot}$
pc$^{-2}$. IC~10 has a substantial molecular gas content compared to
other galaxies of its size, suggesting that the period of vigorous
star formation may be ongoing. IC~10's CO luminosity of $\sim 2 \times
10^{6}$ K km s$^{-1}$ pc$^2$ is an order of magnitude higher than that
of the SMC \citep[$\sim 8 \times 10^{4}$ K km s$^{-1}$
pc$^2$,][]{MIZUNO01} or NGC~1569 \citep[$\sim 10^{5}$ K km s$^{-1}$
pc$^2$,][]{GREVE96,TAYLOR99}. Indeed, \citet[][]{WILCOTS98} calculated
only $\sim 7$ -- $21$ supernovae would be needed to create all of
dramatic holes in the \hi\ distribution; thus, based on the observed
number of WR stars --- possibly as high as 100 and each representing a
future supernova --- IC~10 may have only experienced a small fraction
of the Type II supernovae in store for it in the near future. This
would seem to place IC~10 in contrast to the comparatively quiescent
SMC and the post starburst NGC~1569 \citep[where a burst of star
formation may have ended as recently at 5 Myr ago][]{GREGGIO98}.

Figures \ref{IC10HIANDSTARS} and \ref{IC10HALPHA} show maps of the
atomic gas content, stellar surface density, and H$\alpha$ surface
brightness in IC~10, with the positions of spectroscopically confirmed
WR stars \citep[][]{CROWTHER03} noted in Figure \ref{IC10HALPHA}
\citep[the larger, $\sim 100$, number of WR stars from][ await
spectroscopic confirmation and publication of their
locations]{MASSEY02}. The complex \hi\ structure is described by
\citet[][]{WILCOTS98} in the following way: the stellar disk lies
within a larger \hi\ disk which shows rotation aligned with the
stellar disk, the whole galaxy lies within a much more extended \hi\
structure that is counter-rotating and complex. The \hi\ disk extends
beyond the stars to the east of the galaxy and forms part of a
contiguous position-velocity structure with the extended \hi\ envelope
to the south and east of the galaxy. This large envelope may be have
recently interacted with the galaxy in a way that triggered the
present star formation in the galaxy. Most of the star formation lies
in the central part of the disk, where a large \hi\ cloud is the site
of much of the ongoing star formation activity. The dramatic holes
that give the disk its disturbed (almost spiral) morphology are
probably a result of stellar winds or perhaps supernovae.

Because it lies close to the Galactic plane ($b = -3.3$), the distance
to IC~10 remains uncertain. Current estimates of the distance range
from a lower limit of $500$ kpc \citep[][]{SAKAI99} to $950$ kpc
\citep[][]{HUNTER01}. In this paper, we adopt $950$ kpc, the distance
obtained by \citet[][]{HUNTER01} using the tip of the red giant branch
and a reddening confirmed by comparison with color magnitude
diagrams. The resulting distance is close to $1$ Mpc, a value
frequently adopted in the literature, making for easy comparison. The
values presented in Table \ref{IC10PROPS} and the results of this
paper have been scaled to this distance of $950$ kpc.

\begin{deluxetable*}{l l l}
\tablewidth{5in}
\tabletypesize{\small}
\tablecolumns{3}
\tablecaption{\label{IC10PROPS} Properties of IC~10}

\tablehead{ \colhead{Property} & \colhead{Value} & \colhead{Reference}}

\startdata
Hubble Type & dIrr & \citet[][]{MATEO98} \\
Distance & 950 kpc & \citet[][]{HUNTER01} \\
Axis Ratio & 0.67 & \citet[][]{2MASSLGA} \\
Dynamical Mass & $\sim 1.7 \times 10^{9}$ M$_{\odot}$ &
\citet[][]{MATEO98} \\
Absolute B Magnitude & $-16.3$ \tablenotemark{a,~b}
& \citet[][]{GILDEPAZ03} \\
Stellar Mass & $4 \times 10^{8}$ M$_{\odot}$ & \citet[][]{2MASSLGA} \\
\hi\ Mass & $2\times10^{8}$ M$_{\odot}$ \tablenotemark{a} &
\citet[][]{HUCHTMEIER88}  \\
H$\alpha$ Luminosity & $1 \times 10^{40}$ erg s$^{-1}$\tablenotemark{~a,~b}& \citet[][]{GILDEPAZ03,THRONSON90} \\
FIR Luminosity & $5 \times 10^7$ L$_{\odot}$ \tablenotemark{a} &
\citet[][]{MELISSE94} \\
1.4 GHz Luminosity & $3.4 \times 10^{19}$ W Hz$^{-1}$\tablenotemark{~a}
& \citet[][]{WHITE92} \\
Metallicity & log O/H + 12 = 8.25 & \citet[][]{LEQUEUX79,GARNETT90} \\
\hline
SFR$_{H\alpha}$ & $0.2$ M$_{\odot}$ yr$^{-1}$\tablenotemark{~c} &
\citet[][]{GILDEPAZ03,KENNICUTT94} \\
SFR$_{FIR}$ & $0.05$ M$_{\odot}$ yr$^{-1}$ &
\citet[][]{MELISSE94,BELL03} \\
SFR$_{1.4~GHz}$ & $0.07$ M$_{\odot}$ yr$^{-1}$ &
\citet[][]{WHITE92,BELL03} \\
$\Sigma_{SFR}$ & $0.1$ M$_{\odot}$ yr$^{-1}$ kpc$^{-2}$\tablenotemark{~d} & \\
\hline
CO Luminosity & & this work (\S \ref{LUMSECT}) \\
\hspace{0.03in} Best Estimate & $2.2 \times 10^{6}$ K km s$^{-1}$
pc$^{2}$ & \\
\hspace{0.03in}  Lower Limit & $1.6 \times 10^{6}$ K km s$^{-1}$
pc$^{2}$ & \\
\hspace{0.03in}  Upper Limit & $2.8 \times 10^{6}$ K km s$^{-1}$
pc$^{2}$ & \\
\hline
\enddata
\tablenotetext{a}{Value has been adjusted to our assumed distance of
950 kpc.}
\tablenotetext{b}{Quantity adjusted to use reddening of $E(B-V) =
  0.77$.}
\tablenotetext{c}{Includes 1.1 magnitudes of internal extinction.}
\tablenotetext{d}{H$\alpha$ SFR divided by optical size.}
\end{deluxetable*}

% DESCRIPTION OF OBSERVATIONS
\section{Observations}

In this paper we present two new datasets: a complete survey of the
disk of IC~10 obtained with the BIMA interferometer \citep[described
by][]{BIMAPAPER}, and single-dish pointed observations obtained with
the Arizona Radio Observatory (ARO) 12m to check for emission missed
by the BIMA survey. In this section we describe the acquisition and
reduction of both data sets and the algorithm we used to identify
signal in the BIMA survey. We also introduce several previously
published data sets that we use in our analysis and summarize our
algorithm for measuring GMC properties.

\subsection{The BIMA Survey}

The BIMA survey was conducted in the most compact BIMA configuration,
the D array, and has a resolution of $14\arcsec$. The survey took
place over the course of four observing seasons: fall 2000, spring
2001, fall 2001, and spring 2002. We observed IC~10 during 34 tracks
ranging from 2 to 13 hours in length for a total of $250$ hours of
observing time. For most tracks, we also observed a planet once or
twice to check the absolute flux of the phase calibrator. Each track
consisted of $\sim 20$ fields arranged on a hexagonal grid with
pointing centers spaced by $78''$ (BIMA has a $\approx100\arcsec$
half-power field of view at 115.27 GHz). The correlator configuration
varied somewhat over the survey (the central velocity and channel
width changed slightly) but a typical setup covered a bandwidth of 100
MHz ($\approx 260$ km s$^{-1}$) near the velocity of IC~10 with a
velocity resolution of $\approx 1$ km s$^{-1}$. The final survey has a
resolution of 3 km s$^{-1}$ across a 150 km s$^{-1}$ bandwidth near
the velocity of IC~10, so these variations in the correlator setup do
not affect the final data. For the survey $1$~K =
$2.04$~Jy~beam~$^{-1}$.

We reduced the observations using the MIRIAD software
package\footnote{\url{http://www.atnf.csiro.au/computing/software/miriad}}
. We corrected the observations of the phase calibrator and the source
for line length variations (BIMA monitors the electronic path length
to each antenna by periodically sending a signal to the antenna and
back and measuring the phase on return, allowing variations to be
removed from the data during reduction). We flagged data with
shadowing or very high amplitudes, and channels at the edge of the
correlator. We adopted the flux for the phase calibrator shown in
Table \ref{OBSTAB} and self-calibrated on it, assuming that the phase
calibrator was unresolved by our $14''$ beam (we see no evidence of
extended structure in our data). We transfered the gains and phases as
a function of time to the source.

We combined the $(u,v)$ data from all 34 tracks, applied a natural
weighting scheme, and inverted them into a spectral line map with
velocity channels $3$ km s$^{-1}$ wide. We applied a CLEAN algorithm
to each plane of the data cube in order to remove artifacts generated
by incomplete $(u,v)$ coverage. We capped the algorithm at $500$
iterations in each map and did not clean sources less than $2\sigma$
in peak brightness. The final maps cover an 75 square arcminutes at
better than 0.2~K sensitivity with an angular resolution of
$14\arcsec$, a velocity resolution of $3$ km s$^{-1}$, and a velocity
coverage spanning LSR velocities from $-400$ km s$^{-1}$ to $-250$ km
s$^{-1}$.

The flux of our phase calibrator, 0102+584, varied by a factor of 3
over the two year duration of the survey, and as a result, the
amplitude calibration of the survey may be somewhat uncertain. We used
values interpolated from the BIMA calibrator monitoring campaign,
adjusted by as much as $30\%$ based on comparisons between 0102+584
and planets in our own observations. Based on the variation of the
phase calibrator and the scatter within the planet/phase calibrator
fluxes within our own data, we estimate that the flux calibration of
the survey is uncertain by $30\%$. These gain errors may be nonuniform
across the survey because different tracks contribute to different
parts of the map.

\begin{deluxetable*}{l l l}
\tabletypesize{\small}
\tablewidth{0pt}
\tablecolumns{5}
\tablecaption{\label{OBSTAB} The BIMA Survey}

\tablehead{ \colhead{Observing Season} & \colhead{Flux of Calibrator} & 
\colhead{Dates of Tracks (Duration in Hours)}}

\startdata
Fall 2000 (45h) & 2.5 Jy & 11 Oct (10)\tablenotemark{a}, 12 Oct (12)\tablenotemark{a},
16 Oct (10), 22 Oct (13) \\
\\
Spring 2001 (75h) & 1.3 Jy & 7 Jun (10), 10 Jun (10), 12 Jun (11), 16 Jun (10), \\
& & 17 Jun (10), 21 Jun (5), 23 Jun (4), 29 Jun (4)\tablenotemark{b}, \\
& & 1 Jul (3), 3 Jul (8)\tablenotemark{a} \\
\\
Fall 2001 (83h) & 2.0 Jy & 23 Sep (7), 26 Sep (13), 28 Sep (6), 2 Oct (6), \\
& & 3 Oct (11), 6 Oct (5), 7 Oct (5), 12 Oct (8), \\
& & 13 Oct (5), 20 Oct (8), 21 Oct (9) \\
\\
Spring 2002 (71h) & 3.0 Jy & 25 May (8), 26 May (10), 27 May (5), 28 May (8)\tablenotemark{a}, \\
& & 29 May (3), 31 May (8), 6 Jun (9), 8 Jun (2), \\
& & 13 Jun (7), 15 Jun (5), 16 Jun (6) \\
\enddata
\tablenotetext{a}{Data not used.}
\tablenotetext{b}{Track used different calibrator (0228+673).}
\end{deluxetable*}

\subsubsection{Sensitivity of the Survey}

On average, $50\%$ of the time in a given track was spent integrating
on IC~10. The remainder of the time was spent on calibration and
slewing. The $250$ hours of observing time used for the survey
translates into $\approx 125$ hours of integration on source. The
total area targeted was $13\arcmin\times 8\arcmin\approx100$ square
arcminutes. This corresponds to an integration time per pointing of
$\sim 2$ hours. The theoretical $1\sigma$ noise for $2$ hours of
observation with the D array at $3$ km s$^{-1}$ velocity resolution
and our typical system temperature ($500$ K) is $\sim 0.05$ K. Figure
\ref{SENSMAP} shows a map of the sensitivity of the survey, measured
from the RMS variation in the signal-free channels at the edges of the
map, which is somewhat higher than the estimate above due to missing
antennas, atmospheric decorrelation, and other inefficiencies. About
$40$ square arcminutes of our survey have a $1\sigma$ sensitivity
better than $0.1$~K, another $35$ square arcminutes have $1\sigma$
sensitivities between $0.1$ and $0.2$~K. Thus the final survey covers
$75$ square arcminutes at better than $0.2$~K sensitivity. The noise
in the survey is quite Gaussian. We find $15.9\%$, $2.3\%$, and
$0.13\%$ of the emission in the final cube to lie below $-1\sigma$,
$-2\sigma$, and $-3\sigma$, respectively, almost exactly what would be
expected from a normal distribution.

\begin{figure}
\plotone{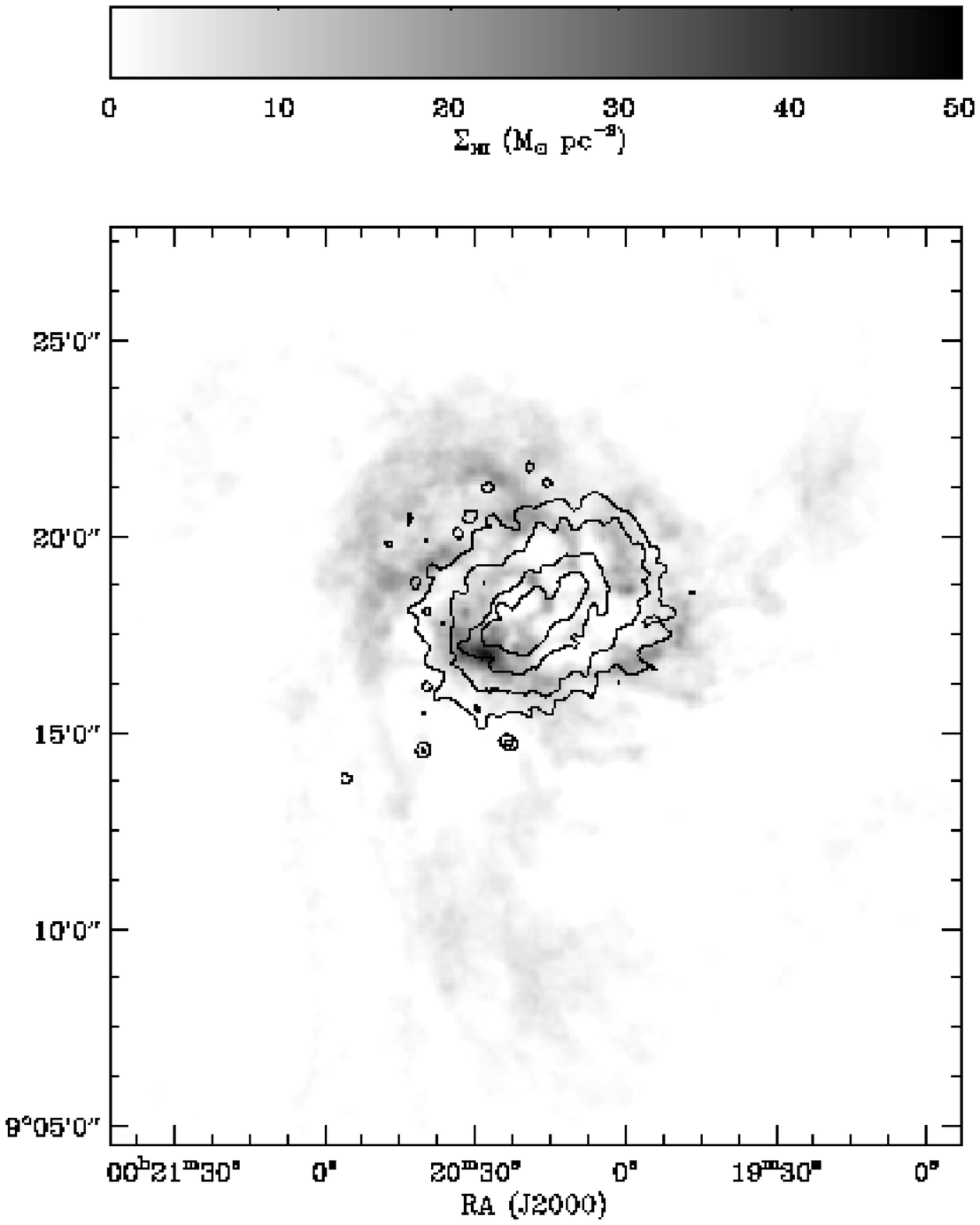}

\figcaption{\label{IC10HIANDSTARS} \hi\ emission from IC~10
\citep[grayscale,][]{WILCOTS98} with stellar surface density contours
overlaid \citep[black, derived from $K$-band data from
][]{2MASSLGA}. The stellar surface density contours are at 50, 100,
200, and 300 M$_{\odot}$ pc$^{-2}$. The \hi\ distribution extends well
beyond the optical disk, stretching for more than $20$ arcminutes
($>5.5$ kpc). \hi\ contours show surface densities of 1, 10, 20, and
30 M$_{\odot}$ pc$^{-2}$. The surface densities shown are line of
sight surface densities, with no correction for inclination.}

\end{figure}

\begin{figure}
\epsscale{1.0} 
\plotone{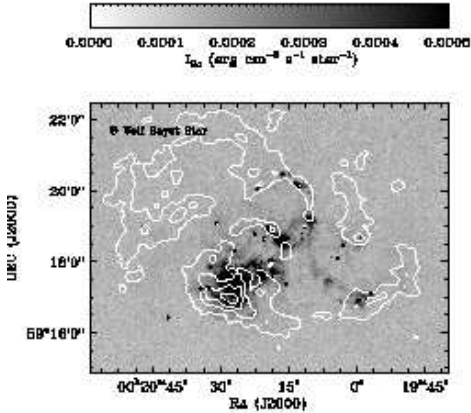}

\figcaption{\label{IC10HALPHA} H$\alpha$ emission from IC~10
\citep[][]{GILDEPAZ03} with \hi\ contours \citep[10, 20, and 30
M$_{\odot}$ pc$^{-2}$, ][]{WILCOTS98} overplotted. The locations of
spectroscopically confirmed WR stars \citep[][]{CROWTHER03} are shown
for comparison.}

\end{figure}

\begin{figure}
\epsscale{1.0} 
\plotone{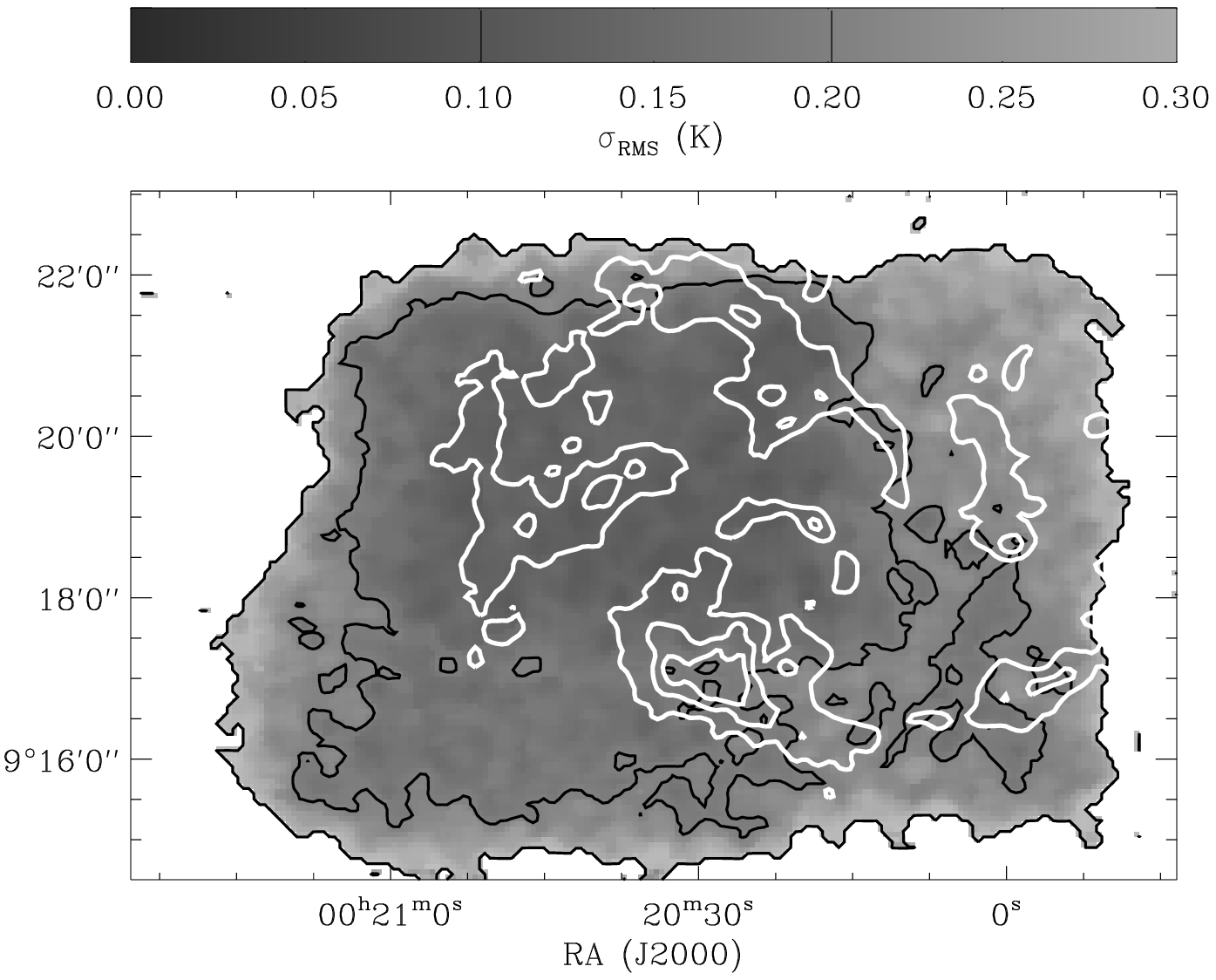}

\figcaption{\label{SENSMAP} Sensitivity of the BIMA D-array survey in
Kelvin. The white contours show \hi\ surface density \citep[10, 20,
and 30 M$_{\odot}$ pc$^{-2}$, ][]{WILCOTS98}. The black contours show
sensitivities of 0.3 and 0.6 K km s$^{-1}$ per 3 km s$^{-1}$ channel.}

\end{figure}

\subsubsection{Signal Identification in the Survey}
\label{SIGNALSECT}

We identified signal using the statistic of merit equal to the the
product of the probability of generating the observed signal or higher
in each of five adjacent velocity channels along a line of sight
(referred to by the subscripts $1$ to $5$) from Gaussian noise. This
statistic, $X$, is calculated via the formula

\begin{equation}
X = P(\frac{I_1}{\sigma}) \times P(\frac{I_2}{\sigma}) \times
P(\frac{I_3}{\sigma}) \times P(\frac{I_4}{\sigma}) \times
P(\frac{I_5}{\sigma})
\end{equation}

\noindent where $P(\frac{I_i}{\sigma})$ is the probability of
generating an intensity equal to or greater than $I_i$ from a
distribution of normally distributed noise with standard deviation
$\sigma$. For Gaussian noise, $P(\frac{I_i}{\sigma})$ is given by

\begin{equation}
P (\frac{I_i}{\sigma}) = \left\{ \begin{array}{ll}
\frac{1}{2} \left[1 - \mbox{erf} \left( \frac{I_i}{\sqrt{2} \sigma}
\right) \right] & \left( I_i > 0 \right)\\ \\
1 & \left( I_i \leq 0 \right) \\
\end{array}   \right. \mbox{ .}
\end{equation}

\noindent The factor of $\frac{1}{2}$ results from normalizing the
error function so that $\mbox{erf} \left(\infty\right)=1$ (rather than
$\frac{1}{2}$). We consider all negative intensities to be the result
of noise and so assign those data a $P (\frac{I_i}{\sigma})$ of
1. Therefore negative intensities can never contribute to a detection
(which are identified by their low values of $X$).

We used this statistic to identify lines of sight with significant
emission in the BIMA survey. We calculated $X$ for all combinations of
5 adjacent velocity channels in the survey. We then select all regions
with $X \leq 10^{-9}$ and RMS sensitivity of $0.2$~K or better. This
value of $X$ corresponds to $\approx 2\sigma$ emission across 5
channels, a single channel containing $\approx 6 \sigma$ emission, or
a range of intermediate cases.  We chose this value, $X \leq 10^{-9}$,
so that we do not expect a false detection over the BIMA survey. We
checked our expected false positive rate by applying this algorithm to
the negative part of the data set (which should consist only of
noise). The algorithm identifies no significant emission in the
negative part of the data cube.

\begin{figure*}
\epsscale{1.0}
\plotone{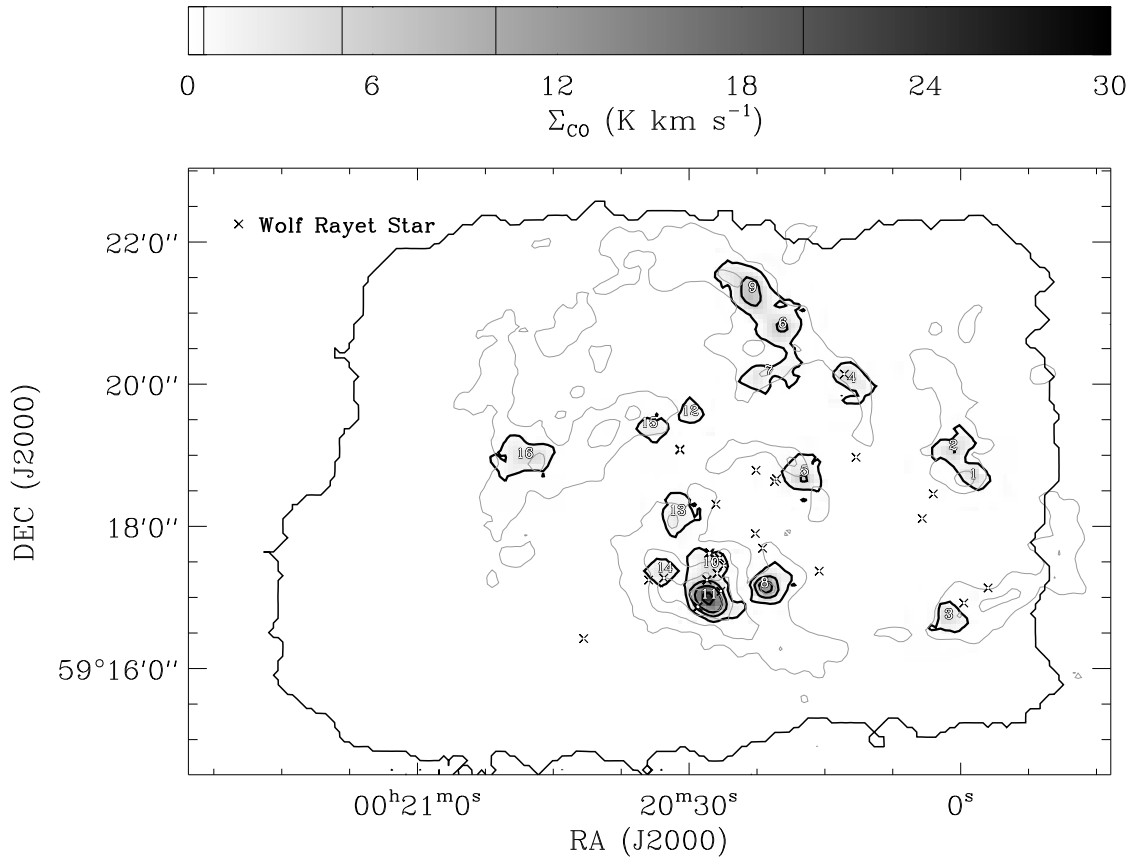}

\figcaption{\label{COMAP} The integrated intensity map resulting from
the BIMA D array survey (grayscale). Black contours show the 1, 10,
20, and 30 K km s$^{-1}$ surface brightness contours. Numbers indicate
the corresponding entry in Table \ref{DRAPROPTAB}. Light contours
indicate \hi\ column densities of 10, 20, and 30 M$_{\odot}$ pc$^{-2}$
\citep[][]{WILCOTS98} and the locations of spectroscopically confirmed
WR stars are indicated \citep[][]{CROWTHER03}. CO emission is almost
exclusively found along lines of sight with $\Sigma_{HI} > 10$~
M$_{\odot}$ pc$^{-2}$ ($N(\hi) = 1.25 \times 10^{21}$ cm$^{-2}$).}
\end{figure*}

We constructed a mask from the signal we identified. The mask is a set
of flags (ones and zeros) that identifies regions of significant
emission in the data cube (a 1 for a region containing emission, a 0
for a signal free region). In the analysis below we use the emission
measured in the BIMA survey multiplied by the mask, so that signal
free regions are treated as having zero emission while the data is
used in regions containing significant signal. To be included in the
mask, we required that region of the survey to have $X \leq 10^{-9}$
across most of a beam (at least three adjacent quarter-beam-sized
pixels). In order to ensure that we did not miss extended, low-lying
emission around the CO peaks, we included all regions within three
velocity channels (a typical cloud line width) and one and a half
beams of each significant region. As a result, the mask contains some
noise at the edge but it is less likely to miss extended emission. The
integrated intensity from the survey multiplied by the mask is shown
in Figure \ref{COMAP}.

This method does not produce a single point source sensitivity, but in
worst case --- $5$ channels at $2\sigma$ significance in the region of
$0.2$ K sensitivity --- we find all emission with an integrated
intensity above 6 K km s$^{-1}$. This corresponds to $1.4 \times 10^5$
M$_{\odot}$ over a BIMA synthesized beam at a Galactic CO-to-H$_2$
conversion factor (appropriate for the IC~10 GMCs but perhaps not for
the diffuse gas). We detect emission with narrow line widths in high
sensitivity regions down to even smaller luminosities, corresponding
to masses of $\lesssim 10^5$ M$_{\odot}$ ($1.8$ K km s$^{-1}$, or $4
\times 10^4$ M$_{\odot}$ over a BIMA beam, in the best case). A
typical sensitivity over the whole survey is therefore $10^5$
M$_{\odot}$.

\subsection{ARO 12m Observations}

\begin{figure}
\epsscale{1.0} 
\plotone{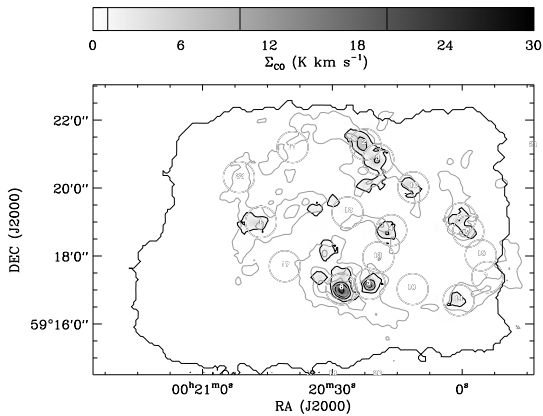}

\figcaption{\label{BIMAAND12M} The integrated intensity map resulting
from the BIMA D-array survey and the locations of the UASO 12m
pointings. The ARO 12m pointings cover most of the CO emission from
IC~10. Light contours indicate $\Sigma_{HI}$ = 10, 20, and 30
M$_{\odot}$ pc$^{-2}$ \citep[][]{WILCOTS98}. Several ARO 12m pointings
are outside this figure, and appear along the edges. The extent of the
survey is indicated by the solid black lines.}

\end{figure}

We observed 22 pointings in IC~10 with the Arizona Radio Observatory
(ARO) 12m telescope at Kitt Peak. This telescope has a $55\arcsec$
($\approx250$~pc) half-power beam-width at 115.27 GHz. The data were
acquired during an observing run in 2002 May 8--16. The locations of
the ARO 12m pointings are listed in Table \ref{12MTABLE} and plotted
along with the results of the BIMA survey and the \hi\ contours in
Figure \ref{BIMAAND12M} --- though note that four of the pointings lie
outside the map. The ARO 12m pointings were selected for two
reasons. First, they were intended to check the flux measurements and
line widths of the BIMA survey --- especially to search for the
presence of extended emission that might be resolved out by the
interferometer. Second, they were selected to provide an independent
search for CO emission at several points of interest not detected by
BIMA --- including \hi\ peaks in the extended emission (possible
locations of CO beyond the optical disk) and two of the \hi\ holes
identified by \citet[][]{WILCOTS98}.

We observed both polarizations either with the 1 MHz filter banks or
with the millimeter autocorrelator. We observed most pointings for
about one hour, though several pointings were bright enough to be
detected in only $\sim30$ minutes. We position switched every six
minutes, with the reference position separated from the pointing by 3
arcminutes in azimuth. In only one case, pointing 14, do we find
evidence of emission in the reference positions close to the velocity
of IC~10. Every six hours, after sunset, and after sunrise, a planet
or other strong continuum source was observed to optimize the pointing
and focus of the telescope. The median system temperature was 345 K.

We reduced the spectrum for each six-minute scan in the following
manner. We removed noise spikes and bad channels by flagging all
channels in each six minute scan with absolute values above the
$5\sigma$ level (none of our sources were nearly this bright in a
single scan). Several channels were known to be bad {\it a priori} and
we flagged these as well. We then subtracted a linear baseline from
the spectrum and binned it to a resolution of $\sim 4$
\kmpers. Finally, we averaged both polarizations and all scans to
produce the final spectrum for each position.

The results of the ARO 12m pointings are summarized in Table
\ref{12MTABLE}. For spectra with detected signal we quote the peak
temperature and the central velocity, velocity width, and integrated
intensity from a three parameter Gaussian fit. For spectra without
detections we quote a $3\sigma$ upper limit to the integrated
intensity assuming a source $15$ km s$^{-1}$ wide (a typical FWHM line
width for our detections). For spectra with detected emission, we also
quote the results from the BIMA and OVRO data convolved to the
resolution of the ARO 12m. We discuss the results of the comparison
among the three datasets in Section \ref{COMPSECTION}. The final
column indicates the BIMA survey clouds associated with each pointing.

\begin{deluxetable*}{l l l l c c c c c c}
\tabletypesize{\tiny}
\tablewidth{0pt}
\tablecolumns{5}
\tablecaption{\label{12MTABLE} ARO 12M Results}

\tablehead{ \colhead{Pointing} & \colhead{Telescope} & \colhead{$\alpha$}
& \colhead{$\delta$} & \colhead{$1\sigma$ Noise\tablenotemark{a}} &
\colhead{$T_{peak}$} & \colhead{$v_{ctr}$} & \colhead{$\sigma_v$} & 
\colhead{$I_{int}$\tablenotemark{b}} & \colhead{Clouds} \\
 & & (J2000) & (J2000) & (mK) & (K) & (km s$^{-1}$) & (km s$^{-1}$) &
(K km s$^{-1}$) & Overlapped }

\startdata
1 & ARO 12m & 00 20 46.4 & 59 18 59.9 & $20$ & $0.10$ & $-335.7$ &
$8.2$ & $1.7$ & B16 \\
1 & BIMA\tablenotemark{c} &  &  & & $0.04$ & $-332.4$ & $4.2$ & $0.4$ & \\
1 & OVRO\tablenotemark{c} &  &  & & $0.03$ & $-332.0$ & $3.0$ & $0.2$
& \\
\\
2 & ARO 12m & 00 20 22.3 & 59 21 16.9 & $19$ & $0.26$ & $-329.6$ & $3.9$ & $2.6$
& B9 \\
2 & BIMA\tablenotemark{c} &  &  & & $0.13$ & $-330.0$ & $3.9$ & $1.3$ \\
2 & OVRO\tablenotemark{c} &  &  & & $0.05$ & $-330.0$ & $4.2$ & $0.5$ \\
\\
3 & ARO 12m & 00 20 19.4 & 59 20 50.4 & $26$ & $0.20$ & $-332.6$ & $5.0$ & $2.5$
 B6\\
3 & BIMA\tablenotemark{c} &  &  & & $0.08$ & $-337.4$ & $4.6$ & $0.9$ \\
3 & OVRO\tablenotemark{c} &  &  & & $0.04$ & $-336.7$ & $3.6$ & $0.4$ \\
\\
4 & ARO 12m & 00 20 16.4 & 59 18 45.5 & $14$ & $0.06$ & $-332.6$ & $7.0$ & $1.0$
& B5 \\
4 & BIMA\tablenotemark{c} &  &  & & $0.04$ & $-337.1$ & $3.8$ & $0.4$ \\
4 & OVRO\tablenotemark{c} &  &  & & $0.04$ & $-333.9$ & $2.6$ & $0.3$ \\ %\vspace{0.05in} \\
\\
5 & ARO 12m & 00 20 21.3 & 59 17 10.6 & $36$ & $0.32$ & $-337.8$ &
$4.5$ & $3.3$ & B8 \\
5 & BIMA\tablenotemark{c} &  &  & & $0.14$ & $-338.7$ & $3.2$ & $1.2$ \\
5 & OVRO\tablenotemark{c} &  &  & & $0.23$ & $-339.4$ & $3.3$ & $2.2$ \\ %\vspace{0.05in} \\
\\
6 & ARO 12m & 00 20 27.9 & 59 17 01.0 & $31$ & $0.31$ & $-330.5$ & $11.1$ &
$8.1$ & B11, B10\tablenotemark{e}  \\
6 & BIMA\tablenotemark{c} &  &  & & $0.20$ & $-330.5$ & $5.2$ & $2.9$ \\
6 & OVRO\tablenotemark{c} &  &  & & $0.20$ & $-330.1$ & $3.3$ & $3.3$ \\ %\vspace{0.05in} \\
\\
7 & ARO 12m & 00 19 54.2 & 59 17 22.6 & $9$ & $0.08$ & $-372.8$ & $2.4$ & $0.6$
& \\
7 & BIMA\tablenotemark{c} &  &  & & $0.06$ & $-373.9$ & $3.7$ & $0.6$ \\
7 & OVRO\tablenotemark{c} &  &  & & $0.07$ & $-371.9$ & $1.6$ & $0.3$
\\ %\vspace{0.05in} \\
\\
8 & ARO 12m & 00 19 58.6 & 59 18 43.2 & $24$ & $0.10$ & $-366.4$ & $3.6$ &
$1.1$ & B1, B2\tablenotemark{e} \\
8 & BIMA\tablenotemark{c} &  &  & & $0.06$ & $-368.4$ & $3.7$ & $0.6$ \\
8 & OVRO\tablenotemark{c} &  &  & & $0.03$ & $-366.5$ & $2.6$ & $0.2$ \\ %\vspace{0.05in} \\
\\
9 & ARO 12m & 00 20 00.3 & 59 19 05.9 & $15$ & $0.12$ & $-368.5$ & $4.0$ &
$1.1$ & B2, B1\tablenotemark{e} \\
9 & BIMA\tablenotemark{c} &  &  & & $0.07$ & $-368.5$ & $3.9$ & $0.7$ \\
9 & OVRO\tablenotemark{c} &  &  & & $0.03$ & $-366.2$ & $2.0$ & $0.2$ \\ %\vspace{0.05in} \\
\\
10 & ARO 12m & 00 20 11.3 & 59 20 01.3 & $17$ & $0.08$ & $-343.1$ & $3.4$ &
$0.7$ & B4 \\
10 & BIMA\tablenotemark{c} &  &  & & $0.07$ & $-346.0$ & $3.6$ & $0.6$ \\
10 & OVRO\tablenotemark{c} &  & & & $0.03$ & $-343.3$ & $1.9$ & $0.2$ \\ %\vspace{0.05in} \\
\\
11 & ARO 12m & 00 20 39.2 & 59 21 12.1 & $19$ & \nodata & \nodata & \nodata &  $\leq 0.51$ \\

12 & ARO 12m & 00 20 26.5 & 59 19 16.7 & $18$ & \nodata & \nodata & \nodata &
$\leq 0.49$ \\

13 & ARO 12m & 00 20 19.4 & 59 17 57.5 & $17$ & \nodata & \nodata &
\nodata &  $\leq 0.46$\\

\\
14 & ARO 12m & 00 20 00.8 & 59 16 41.8 & $10$ & \nodata & \nodata & \nodata &
$\leq 0.27$\tablenotemark{d} & B3 \\
14 & BIMA\tablenotemark{c} & & & & $0.07$ & $-364.7$ & $1.8$ & $0.5$ \\
14 & OVRO\tablenotemark{c} & & & & \nodata & \nodata & \nodata & \nodata \\
\\
15 & ARO 12m & 00 19 55.6 & 59 18 01.1 & $11$ & \nodata & \nodata & \nodata &
$\leq 0.30$\\

16 & ARO 12m & 00 20 11.4 & 59 17 01.5 & $12$ & \nodata & \nodata & \nodata &  $\leq 0.32$\\

17 & ARO 12m & 00 20 40.9 & 59 17 41.9 & $14$ & \nodata & \nodata & \nodata &  $\leq 0.38$\\

18 & ARO 12m & 00 20 30.0 & 59 13 00.0 & $20$ & \nodata & \nodata & \nodata &  $\leq 0.54$ \\

19 & ARO 12m & 00 20 35.0 & 59 09 42.4 & $21$ & \nodata & \nodata & \nodata &  $\leq 0.57$\\

20 & ARO 12m & 00 20 19.4 & 59 08 35.5 & $13$ & \nodata & \nodata & \nodata &  $\leq 0.35$\\

21 & ARO 12m & 00 19 21.2 & 59 21 13.8 & $15$ & \nodata & \nodata & \nodata &  $\leq 0.40$\\

22 & ARO 12m & 00 20 51.5 & 59 20 19.3 & $13$ & \nodata &\nodata & \nodata & $\leq 0.35$ \\
\enddata 

\tablenotetext{a}{In a $5.2$ km s$^{-1}$ channel. For spectra with a
different channel width the noise is adjusted to this channel width.}

\tablenotetext{b}{Upper limits are 3$\sigma$ upper limits for a source
with velocity width $15$ km s$^{-1}$.}

\tablenotetext{c}{All BIMA and OVRO data have been convolved to match
the resolution of the ARO 12m.}

\tablenotetext{d}{Emission in the off position.}

\tablenotetext{e}{Cloud not wholly within the beam but may contribute
to the ARO pointing.}

\end{deluxetable*}

\subsection{Other Data Used in This Paper}

\subsubsection{OVRO High Resolution CO Maps}

We use the high resolution OVRO observations of IC~10 presented by
\citet[][]{WALTER03} to study spatially resolved GMCs and as a third
point of comparison for CO fluxes and line widths. These data have an
angular resolution (FWHM) of $3.4\arcsec$ ($13.5$ pc) or $4.9\arcsec$
($19.5$ pc), depending on the location in the galaxy, and a velocity
resolution of $\approx 0.65$ km s$^{-1}$. The high resolution and good
signal-to-noise ratio in these data allow us to measure the properties
of individual GMCs. The OVRO primary beam is $\sim 50\arcsec$ at 115
GHz, so these data are not sensitive at all to spatial scales more
extended than $\sim 25\arcsec$.

\subsubsection{H$\alpha$ Imaging}

We use an H$\alpha$ image of IC~10 from \citet[][]{GILDEPAZ03} to
measure the star formation per unit area ($\Sigma_{SFR}$) across
IC~10. We correct the H$\alpha$ intensity for the effects of
extinction and then convert to $\Sigma_{SFR}$ using the calibration by
\citet[][]{KENNICUTT94},

\begin{equation}
SFR~\left[ \frac{\mbox{M}_{\odot}}{\mbox{yr}}\right] = 
\frac{L(H\alpha)}{1.26 \times 10^{41}~\mbox{erg s}^{-1}}~\mbox{,}
\end{equation}
which assumes a Salpeter IMF (from 0.1 to 100 M$_{\odot}$).

Because IC~10 lies near the Galactic plane, foreground extinction is
important. We assume a reddening towards IC~10 of $E(B-V) = 0.77$
\citep[][]{MASSEY95,HUNTER01}, which corresponds to an $R$-band
extinction of $\approx 2$ magnitudes for a Galactic extinction law
\citep[][]{CARDELLI89}. This extinction is consistent with the
Galactic dust column near IC~10 in each direction in the dust map of
\citet[][]{DUSTMAP}. Our value is lower than the extinction adopted by
\citet[][]{GILDEPAZ03}, who derive their extinction from the
\citet[][]{DUSTMAP} pointing directly towards IC~10. That pointing may
be confused by FIR emission from the galaxy itself.

We adjust the H$\alpha$ fluxes from IC~10 to account for 1.1
magnitudes of internal extinction but this is a large source of
uncertainty. The reddening extinction towards unembedded stars is due
overwhelmingly to Galactic dust \citep[][]{HUNTER01}. However,
\citet[][]{YANG93} and \citet[][]{BORISSOVA00} compare radio continuum
and Br$\gamma$ fluxes to H$\alpha$ and find additional reddenings of
$E(B-V) \sim 1$ (total $E(B-V) \sim 1.5$ -- $2$) towards embedded HII
regions, implying extinctions as high as $A_R \sim 2.6$ towards these
sources. Most of the H$\alpha$ emission from IC~10 comes from regions
that are at least somewhat embedded (see Figure \ref{IC10HALPHA}):
80\% comes from lines of sight with \hi\ columns above 10 M$_{\odot}$
pc$^{-2}$ ($1.25 \times 10^{21}$ cm$^{-2}$) and the mean \hi\ column
associated with a bit of H$\alpha$ emission (~$\sum I(H\alpha) N(\hi)
/ \sum I(H\alpha)$~) is $15$ M$_{\odot}$ pc$^{-2}$. In the Milky Way,
this column of gas (assuming it all lies between the observer and the
H$\alpha$) implies a reddening of $E(B-V) \sim 0.4$
\citep[][]{BOHLIN78} and a corresponding $R$-band extinction of $\sim
1.0$ magnitudes. Of course, the \hi\ column is unlikely to all lie in
front of the H$\alpha$ emission, but there is a contribution from dust
associated with molecular gas, as well. We use the H$\alpha$ map to
test the applicability of a ``Schmidt Law'' to IC~10 and this value is
similar to the 1.1 magnitudes of internal extinction assumed by
\citet[][]{KENNICUTT98}, so we adopt that value (1.1) for consistency
but note it to be uncertain by $\sim 1$ magnitude.

\subsubsection{VLA High Resolution HI Map}

We use the high resolution VLA \hi\ map of \citet[][]{WILCOTS98},
reduced using natural weighting, which has a resolution of $11\arcsec$
(44 pc). The VLA map contains mostly data from intermediate arrays (B
and C, minimum baseline $\sim 35$ m, with just 10 minutes of D array
data) and is therefore not sensitive to spatial scales $\gtrsim
15$\arcmin (although power on scales as small as $\sim 7.5$\arcmin may
be attenuated by $\sim 40\%$). In this work, we are interested in the
\hi\ over the central disk (where the CO is). In this region the
structure of the \hi\ varies on $\sim 1$\arcmin\ scales (the holes and
filaments in Figure \ref{IC10HIANDSTARS}) and we do not expect
significant loss of power on these scales. However, the \hi\
surrounding IC~10 is very extended and spatial filtering is important
to large scale studies. For example, \citet[][]{WILCOTS98} find only
60\% of the flux recovered by \citet[][]{SHOSTAK89} using the WSRT. In
turn, \citet[][]{SHOSTAK89} recover only half of the value found by
single dish telescopes. The extended emission being resolved out by
the interferometers has low column density \citep[$\sim1 \times
10^{20}$ cm$^{-2}$][]{WILCOTS98}, a factor of $\sim 10$ lower than the
typical columns associated with \co\ detection \citep[see][for a map
of extended emission missed by the WSRT]{SHOSTAK89}. The large flux
discrepancies come from the large spatial extent of the \hi\ and
should only represent a $\sim 10\%$ correction on the compact, high
column structures associated with the \co.

\subsubsection{$K$-band Photometry}

IC~10 is part of the 2MASS Large Galaxy Atlas \citep[][]{2MASSLGA}. We
use the $K$-band image to trace the stellar population in
IC~10. Because IC~10 is at a low Galactic latitude ($b =
-3.3^{\circ}$) foreground stars represent a serious source of
contamination.  In order to remove these foreground stars, we mask out
the brightest pixels in the 2MASS image.  Our threshold for
identifying ``bright pixels'' corresponds to a stellar surface density
of $1800$ M$_{\odot}$ pc$^{-2}$. We removed all pixels above this
value from the image, as well as all data adjacent to these pixels
with a surface density of $\approx 600$ M$_{\odot}$ pc$^{-2}$ (to
ensure that we clipped the tail of the point spread function for the
stars we remove). The highest stellar surface density we find for the
disk of IC~10 is $\lesssim 500$ M$_{\odot}$, so the galaxy should be
unaffected by the masking. After removing bright pixels, we apply a
$15''$ median filter to the whole image, using the median values to
replace the data we removed. Finally, we adopt a $K$-band mass to
light ratio of $0.5$ M$_{\odot}$ / L$_{\odot,K}$, consistent with the
results found by \citet[][]{SIMON05} in their study of rotation curves
of dwarf galaxies. The scatter they find in their mass-to-light ratios
is $\approx 50\%$ (ranging from $0.3$ to $0.7$ M$_{\odot}$ /
L$_{\odot,K}$) so the stellar surface density is uncertain by the same
amount. The resulting $K$-band surface density map has a resolution
comparable to the BIMA survey (because of the filtering) and is
largely uncontaminated by bright foreground stars.  Figure
\ref{IC10HIANDSTARS} shows this map as contours plotted over the
extended \hi\ distribution.

\subsection{Overview of GMC Property Measurements}
\label{CPROPSECT}

This section summarizes the algorithm described in detail by
\citet[][]{ROSOLOWSKY05}, which we use to measure resolved GMC
properties and correct these measurements for biases
introduced by our limited resolution and sensitivity. We apply this
algorithm to the BIMA survey and to the high resolution OVRO data sets
from \citet[][]{WALTER03} to produce the cloud property measurements
in Tables \ref{DRAPROPTAB} and \ref{CPROPSTAB}.

First, we construct a mask containing all high significance signal in
the data cubes. For the BIMA survey we use the mask generated as
described in \S\ref{SIGNALSECT} and simply assign emission to the
nearest local maximum (see the cloud assignments in Figure
\ref{COMAP}). For the OVRO data, we include all regions with two
adjacent velocity channels both containing emission above $4\sigma$
intensity. We expand the mask to include all emission with two
adjacent velocity channels above $2\sigma$ that is contiguous with the
$4\sigma$ peaks. We then identify significant, independent local
maxima within each cloud. Here ``significant'' means that the maxima
are at a significantly higher intensity ($2\sigma$ greater) than
either the edge of the cloud or the highest isosurface shared with
other local maxima. An ``independent'' maximum is separated from all
other maxima by at least a velocity channel or a full beam width.

From the emission uniquely associated with each maximum (i.e. within
the lowest isosurface containing only that maximum), we measure the
size, line width, and luminosity for that cloud. We make the
measurements using intensity-weighted moment methods (i.e. we measure
spatial and velocity dispersions). From the measured dispersions, we
calculate the radius of the cloud using the definition of
\citet[][]{SOLOMON87}. The line width is the full-width at
half-maximum of the integrated spectrum of the cloud. The luminosity
is the integrated emission from the cloud. We correct these
measurements for biases due to the finite sensitivity and resolution
of the astronomical data. We correct for the finite sensitivity by
extrapolating the measured properties to those we would expect for a
data set with perfect sensitivity (by fitting each property as a
function of boundary isosurface value and extrapolating to a boundary
of $0$~K). We correct for the effects of beam convolution on the
measured size of the GMCs by deconvolving the beam size from the
measured size in quadrature (separately for the major and minor
axes). Since the detailed description of the algorithm and its
characterization is beyond the scope of this paper, we refer the reader
to \citet[][]{ROSOLOWSKY05}.

\begin{deluxetable*}{l l l c c c c}
\tabletypesize{\tiny}
\tablewidth{0pt}
\tablecolumns{5}
\tablecaption{\label{DRAPROPTAB} Cloud Properties from the BIMA Survey}

\tablehead{ \colhead{Cloud \#} & \colhead{$\alpha$} & \colhead{$\delta$}
& $v_{ctr}$ & \colhead{$V_{FWHM}$} &
\colhead{Luminosity} & \colhead{$M_{Lum}$} \\
 & (J2000) & (J2000) & (km s$^{-1}$) & (km s$^{-1}$) & 
($10^3$ K km s$^{-1}$ pc$^2$) & ($10^3$ M$_{\odot}$) \\ 
(1) & (2) & (3) & (4) & (5) & (6) & (7) }

\startdata

B1 &  0h 19m 58.6s & 59$^{\circ}$ 18$'$ 40.6$"$ & -367.4 &   7.7 &   30. &  131. \\
B2 &  0h 20m  0.8s & 59$^{\circ}$ 19$'$  4.9$"$ & -366.3 &  12.4 &   38. &  164. \\
B3 &  0h 20m  1.3s & 59$^{\circ}$ 16$'$ 42.3$"$ & -364.0 &  16.1 &   49. &  213. \\
B4 &  0h 20m 11.9s & 59$^{\circ}$ 20$'$  2.2$"$ & -343.6 &  13.6 &   51. &  221. \\
B5 &  0h 20m 17.2s & 59$^{\circ}$ 18$'$ 43.3$"$ & -335.9 &  17.4 &   50. &  218. \\
B6 &  0h 20m 19.6s & 59$^{\circ}$ 20$'$ 48.1$"$ & -337.8 &  10.5 &   75. &  326. \\
B7 &  0h 20m 21.2s & 59$^{\circ}$ 20$'$  8.1$"$ & -330.6 &  11.4 &   41. &  180. \\
B8 &  0h 20m 21.6s & 59$^{\circ}$ 17$'$  8.8$"$ & -339.7 &  10.5 &  102. &  443. \\
B9 &  0h 20m 22.9s & 59$^{\circ}$ 21$'$ 18.3$"$ & -329.6 &   7.5 &  122. &  533. \\
B10 &  0h 20m 27.6s & 59$^{\circ}$ 17$'$ 26.2$"$ & -329.6 &  15.3 &   57. &  246. \\
B11 &  0h 20m 27.7s & 59$^{\circ}$ 16$'$ 59.4$"$ & -330.9 &  14.9 &  238. & 1036. \\
B12 &  0h 20m 29.8s & 59$^{\circ}$ 19$'$ 34.0$"$ & -326.4 &   4.8 &   12. &   52. \\
B13 &  0h 20m 31.3s & 59$^{\circ}$ 18$'$  9.9$"$ & -324.5 &  12.0 &   25. &  110. \\
B14 &  0h 20m 32.6s & 59$^{\circ}$ 17$'$ 21.6$"$ & -311.1 &  15.0 &   11. &   48. \\
B15 &  0h 20m 34.4s & 59$^{\circ}$ 19$'$ 24.0$"$ & -317.7 &  21.4 &   17. &   76. \\
B16 &  0h 20m 48.1s & 59$^{\circ}$ 18$'$ 58.1$"$ & -330.5 &  13.9 &   61. &  266. \\

\enddata 

\end{deluxetable*}

\begin{deluxetable*}{l l l c c c c c c c}
\tabletypesize{\tiny}
\tablewidth{0pt}
\tablecolumns{5}
\tablecaption{\label{CPROPSTAB} GMC Properties in IC~10}

\tablehead{ \colhead{Cloud} & \colhead{$\alpha$} & \colhead{$\delta$}
& $v_{ctr}$ & \colhead{Radius} & \colhead{$V_{FWHM}$} &
\colhead{Luminosity} & \colhead{$\log_{10}~M_{Lum}$} & 
\colhead{$\log_{10}~M_{vir}$} & \colhead{$M_{Vir}$/$M_{Lum}$} \\
 & (J2000) & (J2000) & (km s$^{-1}$) & (pc) & (km s$^{-1}$) & 
($10^3$ K km s$^{-1}$ pc$^2$) & (M$_{\odot}$) & (M$_{\odot}$) & \\ 
(1) & (2) & (3) & (4) & (5) & (6) & (7) & (8) & (9) & (10) }

\startdata

B1 &  0h 19m 58.6s & 59$^{\circ}$ 18$'$ 40.3$"$ & -366.6 &  11.8 $\pm$   6.1 &   4.0 $\pm$   1.0 &  13.4 $\pm$   2.8 &   4.8 $\pm$   0.1 &   4.6 $\pm$   0.2 & 0.6 $\pm$ 0.5 \\
B2 &  0h 20m  0.9s & 59$^{\circ}$ 19$'$  2.2$"$ & -367.4 &  22.8 $\pm$   9.6 &   5.7 $\pm$   2.0 &   9.6 $\pm$   7.0 &   4.6 $\pm$   0.2 &   5.1 $\pm$   0.3 & 3.3 $\pm$ 3.8 \\
B4 &  0h 20m 12.0s & 59$^{\circ}$ 20$'$  2.5$"$ & -342.7 &  29.4 $\pm$   7.2 &   5.6 $\pm$   1.0 &  20.2 $\pm$   2.7 &   4.9 $\pm$   0.1 &   5.2 $\pm$   0.2 & 2.0 $\pm$ 1.1 \\
B5 &  0h 20m 17.3s & 59$^{\circ}$ 18$'$ 42.0$"$ & -334.0 &  26.2 $\pm$   5.5 &   6.2 $\pm$   1.3 &  25.3 $\pm$   4.7 &   5.0 $\pm$   0.1 &   5.3 $\pm$   0.2 & 1.7 $\pm$ 1.0 \\
B6 &  0h 20m 19.6s & 59$^{\circ}$ 20$'$ 48.0$"$ & -338.2 &  17.7 $\pm$   3.9 &   6.0 $\pm$   1.2 &  27.5 $\pm$   5.3 &   5.1 $\pm$   0.1 &   5.1 $\pm$   0.2 & 1.0 $\pm$ 0.5 \\
B8 &  0h 20m 21.7s & 59$^{\circ}$ 17$'$  9.6$"$ & -339.5 &  24.1 $\pm$   2.6 &   6.1 $\pm$   0.6 &  69.5 $\pm$   3.2 &   5.5 $\pm$   0.0 &   5.2 $\pm$   0.1 & 0.6 $\pm$ 0.1 \\
B7 &  0h 20m 22.1s & 59$^{\circ}$ 20$'$  4.9$"$ & -329.3 &  18.3 $\pm$   6.3 &   4.0 $\pm$   1.1 &  18.0 $\pm$   5.4 &   4.9 $\pm$   0.1 &   4.7 $\pm$   0.2 & 0.7 $\pm$ 0.5 \\
B9a &  0h 20m 22.3s & 59$^{\circ}$ 21$'$  5.6$"$ & -330.5 &  15.9 $\pm$   4.6 &   3.0 $\pm$   0.8 &  14.1 $\pm$   3.2 &   4.8 $\pm$   0.1 &   4.4 $\pm$   0.2 & 0.4 $\pm$ 0.3 \\
B9b &  0h 20m 22.5s & 59$^{\circ}$ 21$'$ 21.4$"$ & -328.7 &  24.5 $\pm$   6.3 &   4.3 $\pm$   1.1 &  18.7 $\pm$   5.3 &   4.9 $\pm$   0.1 &   4.9 $\pm$   0.2 & 1.1 $\pm$ 0.7 \\
B9c &  0h 20m 23.7s & 59$^{\circ}$ 21$'$ 17.9$"$ & -333.8 &  27.2 $\pm$   7.7 &   5.5 $\pm$   2.5 &  22.4 $\pm$   5.8 &   5.0 $\pm$   0.1 &   5.2 $\pm$   0.3 & 1.6 $\pm$ 1.5 \\
B11a &  0h 20m 27.2s & 59$^{\circ}$ 16$'$ 53.8$"$ & -334.0 &  22.1 $\pm$   7.3 &  10.4 $\pm$   2.5 &  76.4 $\pm$  24.1 &   5.5 $\pm$   0.1 &   5.7 $\pm$   0.2 & 1.4 $\pm$ 1.0 \\
B11b &  0h 20m 27.3s & 59$^{\circ}$ 17$'$  5.5$"$ & -331.1 &  14.9 $\pm$   3.0 &   7.4 $\pm$   1.1 &  52.5 $\pm$   8.5 &   5.4 $\pm$   0.1 &   5.2 $\pm$   0.1 & 0.7 $\pm$ 0.3 \\
B11c &  0h 20m 28.1s & 59$^{\circ}$ 16$'$ 57.0$"$ & -325.1 &  17.8 $\pm$   4.5 &  11.2 $\pm$   2.2 &  64.3 $\pm$  12.0 &   5.4 $\pm$   0.1 &   5.6 $\pm$   0.2 & 1.5 $\pm$ 0.8 \\
B11d &  0h 20m 29.0s & 59$^{\circ}$ 17$'$  4.6$"$ & -327.4 &  18.4 $\pm$   3.9 &   7.9 $\pm$   1.5 &  25.6 $\pm$   4.8 &   5.0 $\pm$   0.1 &   5.3 $\pm$   0.2 & 1.9 $\pm$ 1.0 \\

\enddata 

\end{deluxetable*}

% THE RESULTS SECTION
\section{Results}

\subsection{Comparison of the Three CO Datasets}
\label{COMPSECTION}

\begin{figure*}
\epsscale{1.0} 
\plotone{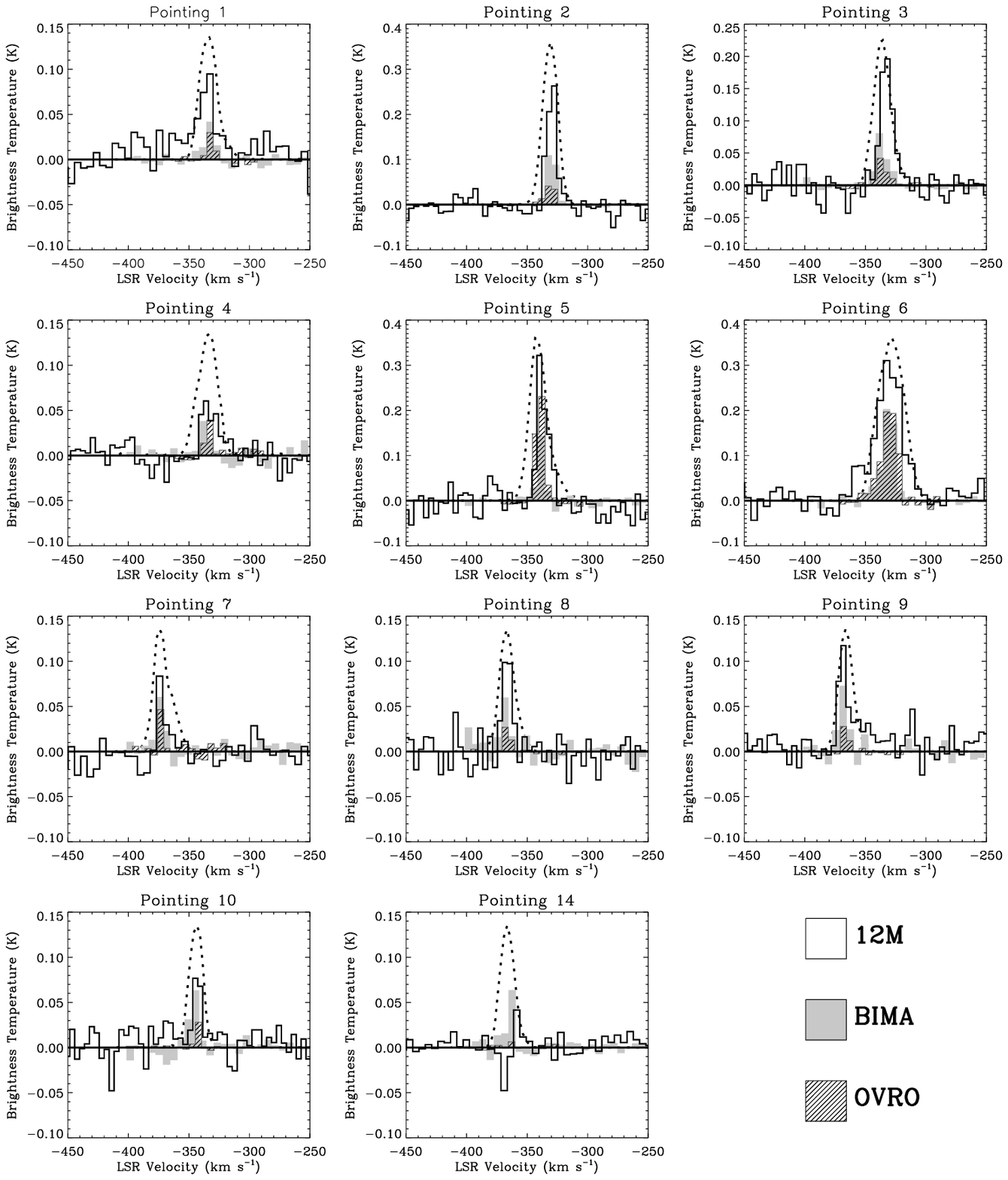}

\figcaption{\label{COMPSPEC} ARO 12m, BIMA, and OVRO spectra compared
for common pointings which show emission. The BIMA and OVRO data have
been convolved to the resolution of the ARO 12m. The \hi\ spectrum,
normalized to 80\% of the $y$-axis range, is overplotted as a dashed
line to indicated the velocity width of the \hi.}

\end{figure*}

How do the emission properties from the BIMA survey, the ARO 12m data,
and the OVRO observations compare? We convolved the BIMA survey, OVRO
data, and the \hi\ data (first clipped at 3$\sigma$) to the resolution
of the ARO 12m data. Figure \ref{COMPSPEC} shows these data for each
ARO 12m pointing with detected emission (the \hi\ spectra are
arbitrarily normalized). Based on these spectra and Figure
\ref{BIMAAND12M} we draw the following conclusions:

\noindent 1. The agreement between the central velocities among all
three datasets is excellent. Further, the velocities are consistent
with the \hi\ emission (dashed line in Figure \ref{COMPSPEC}). Towards
the brightest lines, the width of CO emission detected by the ARO 12m
is comparable to, but always a bit smaller than, that of the \hi\
emission.

\noindent 2. Both interferometers miss emission, and OVRO misses more
than BIMA. This is probably a result of interferometers filtering out
extended emission and not due to low signal-to-noise or to calibration
errors. Both the BIMA survey and the OVRO data are more sensitive than
the ARO 12m (both data sets have RMS noise $<10$~mK over an ARO 12m
beam at $6$~km~s$^{-1}$ velocity resolution, compared to a typical
noise of $15$~mK for the ARO 12m spectra). Further, the OVRO data are
missing more flux than the BIMA survey, despite having a better
signal-to-noise. Finally, the difference is not only a gain offset, as
one would expect from a calibration discrepancy. Rather, the line
widths of the interferometer data (particularly the OVRO data) are
smaller than those found by the single dish, implying that an extended
component with a larger velocity width contributes to the single dish
data but not the interferometer data.

How much emission is resolved out by the interferometers? On average,
BIMA recovers $50\%$ of the integrated intensity found by the ARO 12m,
and OVRO recovers $30\%$. BIMA finds a line width that is on average
$90\%$ of that recovered by the ARO 12m, while OVRO line widths are,
on average, $60\%$ that of the ARO 12m for the same pointing. We
simulated BIMA D array observations of several gaussian sources at the
declination of IC~10. The observations recover 95$\%$ of the flux for
a 20\arcsec\ (FWHM) source, 60\% for a 30\arcsec\ source, 30\% for a
40\arcsec\ source, and 14\% for a 40\arcsec\ source. If the CO
structures in IC~10 are $\sim 35$\arcsec\ (160 pc) we might expect to
lose half of the flux. The actual scales may be slighly more compact
since there are inefficiencies associated with signal identification
and non-ideal observing conditions that will affect diffuse emission
or long baselines.

\noindent 3. We detect CO emission with the ARO 12m only at pointings
which also show emission in the BIMA survey (see point 4). Therefore
the diffuse emission appears to be associated with the larger
GMCs. Further, the ARO 12m pointings cover almost all of the CO
emission found in the BIMA survey. Only $\sim 1 \times 10^5$ K km
s$^{-1}$ pc$^2$ (BIMA clouds 7, 12, 13, 14, and 15) of the emission
from the BIMA survey (about 10\% of the total) lies in clouds not
targeted by the ARO 12m pointings.

\noindent 4. Two pointings deserve specific commentary: numbers 7 and
14. Pointing 7 is detected by the ARO 12m and Figure \ref{COMPSPEC}
shows that it is also detected in the BIMA survey. However, the
emission is not strong enough to be included in our mask. Pointing 14
is detected in the BIMA survey, but the ARO 12m emission appears to be
contaminated by emission in the off position.

Our comparison of the three data sets suggests that $\sim 50\%$ of the
CO emission from IC~10 comes from a high velocity width, spatially
extended component that is resolved out by both OVRO and BIMA. OVRO
further resolves out another 20\% of the CO emission seen by the ARO
12m. Similar results have been found in the Milky Way, M~31 and M~33
\citep[][]{POLK88,BLITZ85,ROSOLOWSKY03}. In those galaxies, too,
diffuse gas, or an extended grouping of small molecular clouds
indistinguishable from diffuse gas, may contribute a large portion of
the CO emission along a line of sight.

\subsection{The Total Content of Molecular Gas and GMC Properties}
\label{LUMSECT}

Figure \ref{COMAP} and Table \ref{DRAPROPTAB} present the results of
the BIMA survey.  The total luminosity from all 10 CO detections with
the ARO 12m of $1.6 \pm 0.3 \times 10^6$ K km s$^{-1}$ pc$^2$ and this
is our formal lower limit to the total CO luminosity of IC~10. If we
stack {\it all} of the ARO pointings and integrate, the luminosity
rises to $2.0 \times 10^6$ K km s$^{-1}$ pc$^2$ with the increase of
25\% due emission not detected in individual pointings \citep[a
signature of a low-intensity diffuse component, perhaps similar to
that found by][]{ISRAEL6822}. The ARO 12m pointings overlap $\sim
90\%$ of the emission found in the BIMA survey, so we estimate that
the total CO luminosity of IC~10 is $\sim 2 \times 10^6 \times
\frac{1}{0.9} = 2.2 \times 10^6$ K km s$^{-1}$ pc$^2$. This number is
quite uncertain because our masking algorithm is chosen to avoid false
positives (rather than for completeness). We estimate an upper limit
by noting that the inclusion of 11 pointings without individual CO
detections raises the integrated luminosity by $\sim 4 \times 10^5$ K
km s$^{-1}$ pc$^2$. The total area in IC~10 with an \hi\ surface
density $> 10$ M$_{\odot}$ pc$^2$ (and therefore likely to harbor
molecular gas) corresponds to $\sim 16$ times the area of the 12m
beam. Of this area about $6$ times the area of the 12m beam is already
covered by our ARO observations. Four of our nondetection pointings
are extremely unlikely sites for molecular gas emission. Therefore, we
might expect another $\frac{10}{7} \times (4 \times 10^5) \sim 6
\times 10^6$ K km s$^{-1}$ pc$^2$ in diffuse emission. We therefore
suggest $2.8 \times 10^6$ K km s$^{-1}$ pc$^2$ as an upper limit to
the CO luminosity. The BIMA survey recovers about half of our best
guess at the luminosity --- the 16 GMCs listed in Table
\ref{DRAPROPTAB} have a total luminosity of $1.0 \times 10^6~ $ K km
s$^{-1}$ pc$^2$ (which includes a $\sim 30\%$ sensitivity correction
as described in \S \ref{CPROPSECT}).

\begin{figure*}
\epsscale{.9} 
\plotone{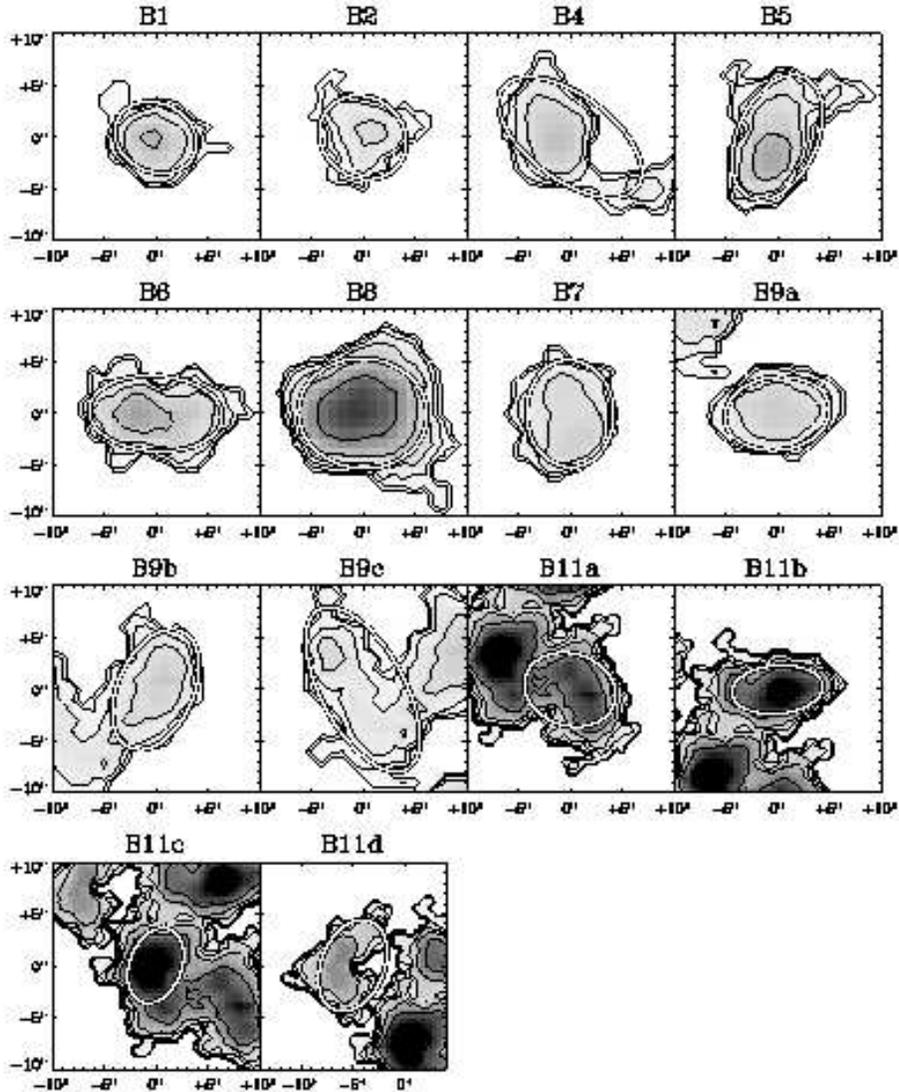}

\figcaption{\label{IC10CLOUDS} Intensity maps of GMCs in IC~10. The
figure shows all CO emission near the center of the GMC. The circles
indicate the measured sizes of the GMC sensitivity or resolution
corrections. At the distance of IC~10 ($950$ kpc), $1\arcsec$
corresponds to 4.6 pc, so the $40\arcsec$ boxes shown have spatial
sizes of $92$ pc.}

\end{figure*}

We used the algorithm of \citet[][]{ROSOLOWSKY05} summarized in
\S\ref{CPROPSECT} to identify clouds and then measure their properties
using the OVRO dataset \citep[][]{WALTER03}. Table \ref{CPROPSTAB}
gives our measurements of the properties of $14$ GMCs. Column (1)
gives the cloud name (which also indicates the BIMA survey cloud to
which the OVRO cloud most nearly corresponds); columns (2) and (3)
give the intensity-weighted average position of the cloud; column (4)
gives the intensity-weighted average velocity of emission from the
cloud; column (5) gives the equivalent spherical radius of the cloud
\citep[][]{SOLOMON87} in parsecs, after correction for the effects of
beam convolution; column (6) gives the FWHM line width of the cloud;
column (7) gives the cloud luminosity in K km s$^{-1}$ pc$^2$; column
(8) gives the mass derived from this luminosity assuming a Galactic
CO-to-H$_2$ conversion factor; column (9) gives the dynamical mass of
the cloud, calculated assuming virial equilibrium and a $\rho \propto
r^{-1}$ density profile, so that $M_{vir} = 189~R~V_{FWHM}^2$, with
$R$ and $V_{FWHM}$ as defined above; column (10) gives the ratio of
virial to luminous mass for the cloud. The virial masses are
comparable to the luminous masses, suggesting that the CO-to-H$_2$
conversion factor does not vary significantly from the Galactic
value. Figure \ref{IC10CLOUDS} shows a map each of cloud. Ellipses
indicate the size of each cloud before any corrections for resolution
or sensitivity effects are applied, so that the sizes shown is
directly comparable to the structures in the data. In Table
\ref{CPROPSTAB} and in the analysis below, however, we use the
corrected values.

We include a brief word of warning regarding cloud names. For clarity
we refer to clouds measured in the OVRO data using the same system we
used for the BIMA survey, using ``a,b,c'' when a BIMA cloud is
resolved into several GMCs by OVRO. However, we have not tried to
systematically identify a single set of clouds using both data sets,
and the association implied by a shared name should not be weighted
too heavily. For example, though no cloud B10 exists in the OVRO data,
emission from that object certainly corresponds to some of the
emission found in clouds B11a,b,c, and d.

\subsection{Comparison to GMC Properties From the Literature}

There are several observations of IC~10 GMCs in the literature: do the
properties we measure agree with these studies?  Table
\ref{GMCCOMPTAB} shows a comparison of the properties of three GMCs in
the bright CO complex in the southeast part of the galaxy (cloud 15 in
Figure \ref{COMAP}, ARO 12m pointing 6 in Figure
\ref{BIMAAND12M}). Table \ref{GMCCOMPTAB} shows the properties
measured by two previous studies \citep[][]{WILSON95,OHTA92} and this
work. All three works decompose the emission in essentially the same
manner, though some of the variations in Table \ref{GMCCOMPTAB} may
arise from decisions about which emission to assign to which GMC. To
properly compare these observations we have adjusted the sizes
measured by the two previous studies to match our definition of the
radius, our adopted distance, and our adopted Galactic CO-to-H$_2$
conversion factor --- both \citet[][]{OHTA92} and \citet[][]{WILSON95}
quote full width half maximum sizes and they adopt distances of 1.3
and 0.82 Mpc, respectively. \citet[][]{OHTA92} labels the GMCs in
question 'NC1,' 'NC2,' and 'SC' (1, 2, and 3 in Table
\ref{GMCCOMPTAB}); \citet[][]{WILSON95} calls them 'MC1,' 'MC2,' and
'MC3;' they are 'B11b,' 'B11c,' and 'B11a,' respectively, in our Table
\ref{CPROPSTAB}. The sizes agree well for two of the three clouds, and
we measure a notably lower size for GMC 1; there is a significant
($\sim 50\%$) scatter in the line widths.  The fluxes measured by
\citet[][]{WILSON95} are systematically lower than what we measure ---
integrated over the complex, \citet[][]{WILSON95} measures half of our
flux. The discrepancy in fluxes and our 15\% larger adopted distance
explain why \citet[][]{WILSON95} derives a higher CO-to-H$_2$
conversion factor for the GMCs than we do. The combination of a larger
adopted distance and the systematically higher fluxes found by
\citet[][]{OHTA92} explain the similarity in the derived CO-to-\htwo\
conversion factors.

Which measurements are closer to the true fluxes? The complex in
question corresponds to pointing 6 with the ARO 12m (see Figures
\ref{BIMAAND12M} and \ref{COMPSPEC} and Table \ref{12MTABLE}) and
cloud BIMA 11 in Table \ref{DRAPROPTAB}. BIMA and OVRO find similar
fluxes for the region (both corresponding to $\approx 10^6$
M$_{\odot}$ for our adopted \xco) and the ARO 12m recovers about 2.5
times this luminosity. \citet{WILSON95} finds fluxes lower than those
found by BIMA, the newer OVRO dataset, or the ARO 12m. The NMA dataset
appears consistent with the BIMA and OVRO results (also finding
$\approx 10^6$ M$_{\odot}$ for our \xco). Therefore the
\citet[][]{WILSON95} fluxes represent an outlier from the other
interferometric data. All four interferometric data sets resolve out a
large fraction of the flux (as measured by the ARO 12m).

\begin{deluxetable*}{l l l l l}
\tabletypesize{\small}
\tablewidth{0pt}
\tablecolumns{5}
\tablecaption{\label{GMCCOMPTAB} IC 10 SE GMC Property Comparison}

\tablehead{\colhead{GMC} & \colhead{Property} & \colhead{Ohta et al. (1992)} &
  \colhead{Wilson (1995)} & \colhead{This Paper}}

\startdata
1 & Radius (pc) & $23$\tablenotemark{a,b,c} & $32$\tablenotemark{a,b,c} & $15$ \\
1 & $V_{FWHM}$ (km s$^{-1}$) & $7.3$ & $10.9$ & $7.4$ \\
1 & Luminous Mass ($10^3$ M$_{\odot}$) & $290$\tablenotemark{a,e} & $190$\tablenotemark{a,e} & $250$ \\
\\
2 & Radius (pc) & $19$\tablenotemark{a,b,c} & $23$\tablenotemark{a,b,c,d} & $18$ \\
2 & $V_{FWHM}$ (km s$^{-1}$) & $7.3$ & $8.9$ & $11.2$ \\
2 & Luminous Mass ($10^3$ M$_{\odot}$) & $230$\tablenotemark{a,e} & $120$\tablenotemark{a,e} & $250$ \\
\\
3 & Radius (pc) & $24$\tablenotemark{a,b,c} & $24$\tablenotemark{a,b,c,d} & $22$ \\
3 & $V_{FWHM}$ (km s$^{-1}$) & $< 8.2$ & $12.8$ & $10.4$ \\
3 & Luminous Mass ($10^3$ M$_{\odot}$) & $490$\tablenotemark{a,e} & $130$\tablenotemark{a,e} & $320$ \\
\hline
\enddata
\tablenotetext{a}{Value has been adjusted to our assumed distance of
950 kpc.}
\tablenotetext{b}{Geometric mean of 2D size from the literature.}
\tablenotetext{c}{Adjusted to our radius definition from FWHM.}
\tablenotetext{d}{Cloud unresolved in one dimension so size is a maximum.}
\tablenotetext{e}{Adjusted to our adopted value of \xco.}
\end{deluxetable*}

\section{Discussion}

\subsection{Molecular Gas Fraction and Depletion Time}

Using a Galactic CO-to-H$_2$ conversion factor, $2 \times 10^{20}$
\xcounits (appropriate for the GMCs but perhaps not for diffuse gas,
see \S\ref{XCOSECT}), the luminosity we estimate for IC~10 translates
to a molecular gas mass of $9 \times 10^{6}$ M$_{\odot}$. This mass is
small compared to the other components of IC~10 --- the stellar, \hi\,
and dynamical masses are $\sim 4,~2,$ and $15 \times 10^8$ M$_{\odot}$
respectively --- and this implies that only $4\%$ of the gas is
molecular. For a star formation rate of $\sim 0.2$ M$_{\odot}$
yr$^{-1}$, the depletion time for the molecular gas associated with CO
in IC~10 is $\sim 4 \times 10^7$ years, compared to the median
depletion time of $2\times10^9$ years found in nearby dwarf galaxies
\citep{LEROY05}. The molecular gas mass is only about $\sim2\%$ of the
stellar mass.  This value tends to be larger, $\sim 15\%$, in LMC-type
dwarf galaxies \citep[][]{YOUNG91,LEROY05}. In fact, the amount of
molecular gas per stellar luminosity tends to be fairly constant among
{\em all} star forming galaxies, large and small \citep[in the
$K$-band this ratio is $\sim0.07$ M$_{\odot}$ /
L$_{K,\odot}$][]{LEROY05}. Although IC~10 is CO-deficient compared to
larger galaxies, it is CO-rich when compared to the SMC or NGC~1569.
\citet[][]{MIZUNO01} find that the CO luminosity of the SMC is
$\approx 8 \times 10^{4}$ K km s$^{-1}$ pc$^2$, more than an order of
magnitude fainter than the CO luminosity of IC~10, while NGC~1569,
which has a CO luminosity of $\sim 10^5$ K km s$^{-1}$ pc$^2$
\citep[][]{GREVE96,TAYLOR99}. This yields an even shorter molecular
gas depletion time for these two systems than what we observe in
IC~10.

\subsection{The CO-to-H$_2$ Conversion Factor in IC~10}
\label{XCOSECT}

With lower abundances of carbon and oxygen, harder radiation fields,
and less dust to shield each parcel of gas, dwarf galaxies may be
expected to display a different relationship between CO emission and
molecular hydrogen content. Calibrating the CO-to-H$_2$ conversion
factor, $X_{CO}$, has often been a goal of CO studies of dwarf
galaxies \citep[e.g.][]{WILSON95}. The key to measuring $X_{CO}$ in
other galaxies is to find an independent method of measuring the
amount of molecular gas present. We attempt an independent estimate of
the mass by measuring the size and line width of a molecular cloud
from its CO emission and then calculating the dynamical mass under the
assumption of virial equilibrium. 

In our analysis of the high resolution CO data, we find that the IC~10
clouds are indistinguishable from GMCs analyzed in the same way in
M~31 and M~33 and that they are very similar to Milky Way clouds. If
the IC~10 clouds are virialized, then the mean CO-to-H$_2$ conversion
factor in the CO peaks is $2.6 \times 10^{20}$ \xcounits\ (if the
clouds are only marginally bound then \xco\ will be half of this
value). Virial mass studies in the Milky Way yield a CO-to-\htwo\
conversion factor $\xco\approx 3 \times 10^{20}$ \xcounits
\citep[][]{SOLOMON87}, similar to the one we measure in IC~10 within
the uncertainties; interferometric studies of M~31 and M~33 yield
approximately the same result
\citet[][]{ROSOLOWSKY03,ROSOLOWSKY05}. Because, in the Milky Way, the
\xco\ value derived from gamma-ray observations is thought by many to
be more reliable (it is independent of the dynamical state of the
GMCs), we adopt the Galactic value of $X_{CO}$ obtained by those
studies, $2 \times 10^{20}$ \xcounits~\citep[][]{GAMMAXCO, DAME01} in
the remainder of this study.

In \S \ref{COMPSECTION} we found that OVRO may resolve out $70\%$ of
the emission. The measurements of the GMC properties are probably
robust, since the GMCs are compact relative to the $\sim100$ pc scales
we expect OVRO to resolve out and the OVRO data have good
sensitivity. However, the GMC properties measured from the OVRO data
do not constrain the CO-to-H$_2$ conversion factor in the extended
emission. One possibility is that the CO resolved out by OVRO and BIMA
comes from a spatially extended collection of small GMCs. Below we
find evidence for a GMC mass spectrum with a power law index of
$-2.0$, which implies that there may be as much mass in GMCs below our
completeness limit as above it. If these low mass GMCs make up the
extended structure that is resolved out by BIMA and OVRO, then we
expect that the CO-to-H$_2$ conversion factor from the OVRO clouds
will apply, at least approximately, to all of the CO emission.

Based on observations of the 158 $\mu$m [CII] emission line,
\citet[][]{MADDEN97} suggested a CO-to-H$_2$ conversion factor much
higher than the Galactic value. They mapped IC~10 at $\sim 1'$
resolution and found that it is luminous in the [CII] line compared to
the CO luminosity. In the northern and western regions, they found
that the minimum amount of hydrogen needed to produce the observed
[CII] emission implies a substantial mass of gas that was not inferred
from the \hi\ or \co\ emission. They suggested that in parts of IC~10
molecular hydrogen may exist under conditions of low extinction, $A_V
\sim 1$ -- $2$, so that CO is dissociated but self-shielded H$_2$ is
abundant. They argue that in these regions the H$_2$ column may exceed
the \hi\ column by a factor of 5. Although we adopt a Galactic \xco,
we can not rule out the evidence of \citet[][]{MADDEN97} for a large
reservoir of non-CO-emitting molecular gas outside the central
GMCs. We do not account for such gas in our discussion of star
formation efficiencies because such gas must be warm, diffuse, poorly
shielded, and, as such, does not seem to be a likely locale for star
formation (the excitation temperature of the 158 $\mu$m [CII] line is
92~K and the CO-free molecular gas posited by \citet[][]{MADDEN97}
exists at extinctions of $A_V \sim 1$ -- $2$ magnitudes). Further, if
such gas exists in other galaxies it will be similarly unaccounted for
by the CO luminosity and should thus be left out for a self-consistent
comparison.

There may be evidence for a reservoir of warm, CO-free H$_2$ beyond
the GMCs, however observations do not suggest the presence of a hidden
reservoir of cold gas. In a study of the dust continuum in the
southeast part of IC~10, \citet[][]{BOLATTO00} considered and rejected
the possibility of a large reservoir of cold molecular gas in that
region. Although they find an excess of long wavelength infrared
emission, they cite the lack of CO self absorption and the normal CO
\jtwo\ to CO \jone\ ratios as evidence that the long wavelength
emission is not indicative of a massive reservoir of cold molecular
gas. \citet[][]{THRONSON90} also finds the amount of 155 $\mu$m dust
emission in IC~10 to be consistent with the \hi\ emission from the
galaxy. A large population of hidden molecular gas is not necessary to
explain their observations, though IC~10 does show a mild excess at
155 $\mu$m.

For the rest of this paper, we adopt a CO-to-H$_2$ conversion factor
of $2 \times 10^{20}$ \xcounits\ for all of the CO emission. We are
confident in the applicability of this value of \xco\ to the GMCs but
less certain whether it is appropriate for diffuse CO emission. We
neglect the possibility of warm H$_2$ untraced by CO because we have
no observational handle on such gas and it seems unlikely to form
stars, but we emphasize that any such excess of molecular gas must
exist outside the most massive GMCs or we would see evidence for it in
the virial masses we measure from the OVRO data. We note the following
conversions for our adopted \xco: an integrated intensity of 1 K km
s$^{-1}$ corresponds to a molecular gas surface density of 4.4
M$_{\odot}$ pc$^{-2}$, including helium, and therefore a luminosity of
1 K km s$^{-1}$ pc$^2$ translates into a molecular mass of 4.4
$M_{\odot}$.

\subsection{Mass-Size-Line Width Relations in IC~10}

In the Milky Way, M~33, and M~31 molecular gas is concentrated in GMCs
which exhibit scaling laws relating their properties
\citep[][]{SOLOMON87,ROSOLOWSKY03,ROSM31}. These relationships, often
referred to as ``Larson's Laws,'' \citep[][]{LARSON81} relate the size
of a GMC to its line width, the CO luminosity to the dynamical mass,
and the luminosity to the line width. In this section we compare
clouds in IC~10 to clouds from the Milky Way, M~33, and M~31. M~31 and
M~33 are at distances comparable to that of IC~10, $\lesssim 1$ Mpc,
and the interferometric data used for the comparison have similar
spatial resolutions ($\approx 20$ pc). We measure their properties
using the same algorithm used to analyze the IC~10 clouds \citep[a
reanalysis of the data presented in][]{ROSOLOWSKY03,ROSM31}. Thus the
M~31 and M~33 data should represent an excellent ``control'' sample,
and any differences between GMC properties in these systems and IC~10
should be a result mainly of environmental factors, not observational
or analytical biases. The Milky Way data have not been analyzed in the
same manner as the other data sets --- they just consist of the GMC
properties measured by \citet[][]{SOLOMON87} --- so systematic
differences may bias the comparison. However, we have attempted to
correct for sensitivity and resolution biases in our data and the
\citet[][]{SOLOMON87} data have good sensitivity and spatial
resolution (being Galactic data) so we anticipate the magnitude of
these biases to be small.

\begin{figure*}
\epsscale{0.45}
\plotone{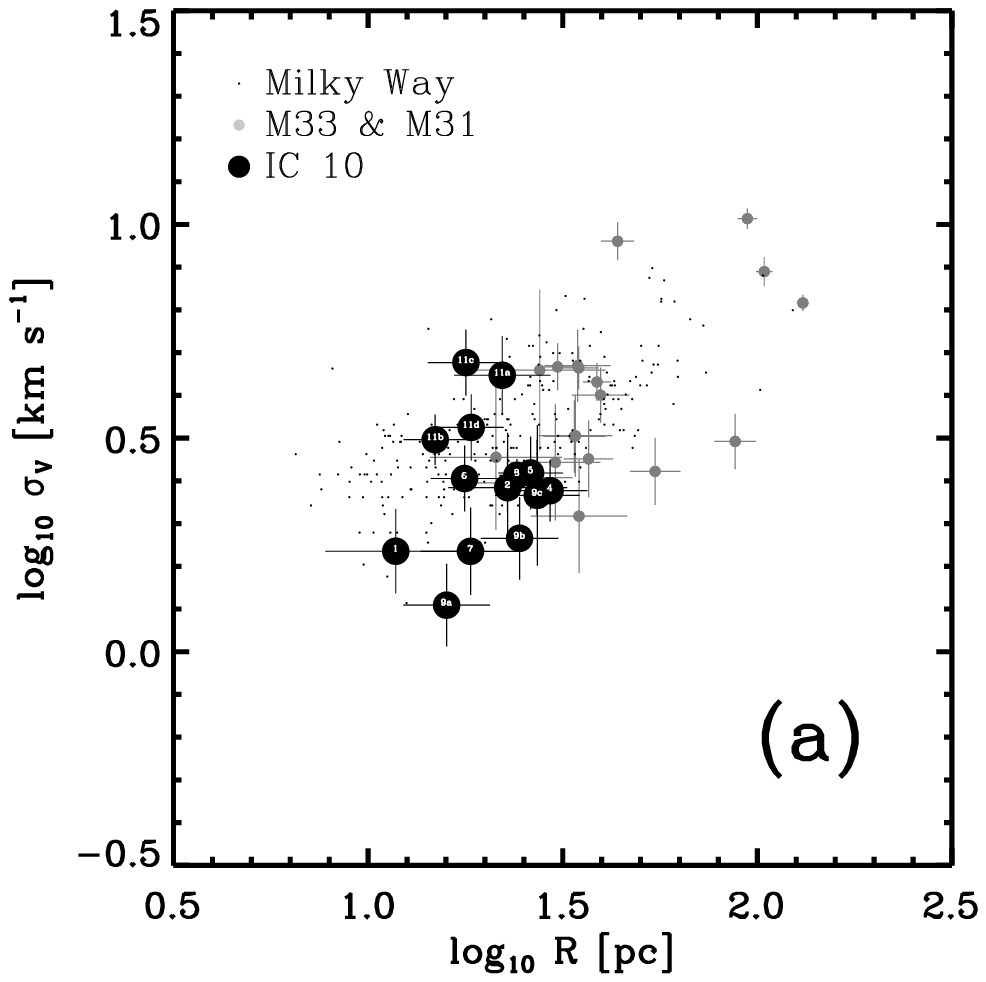}  \plotone{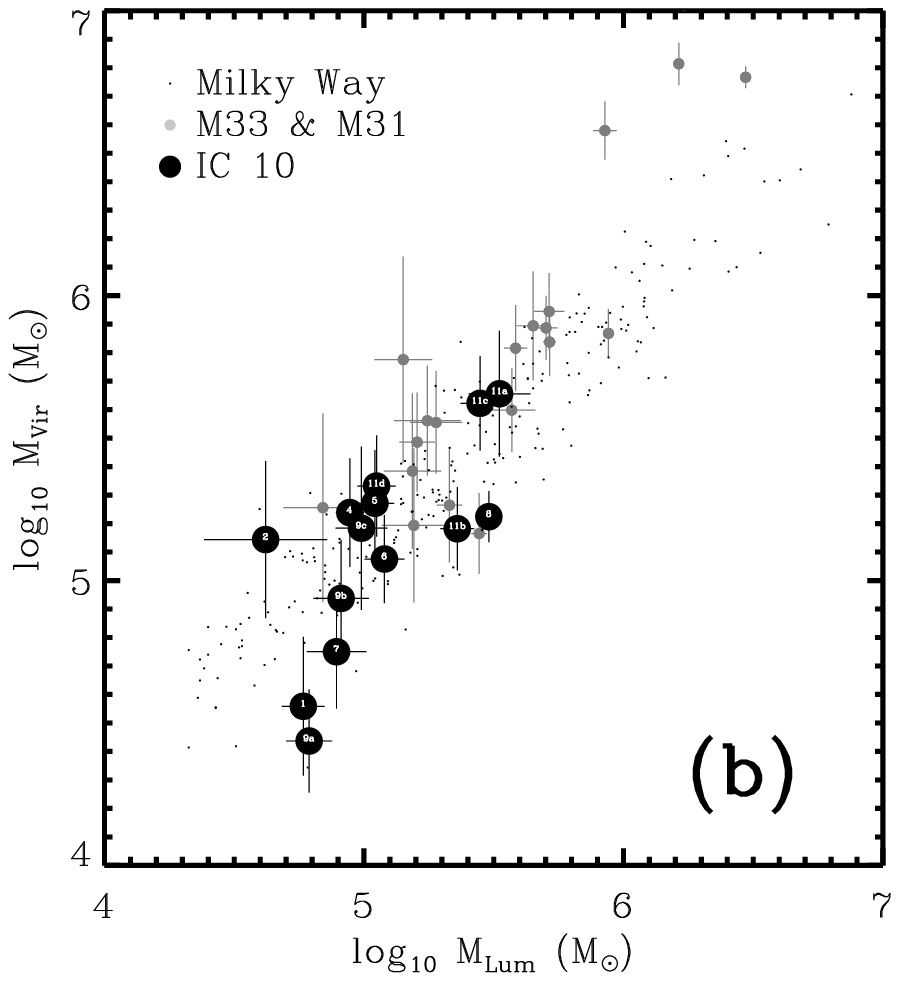} \plotone{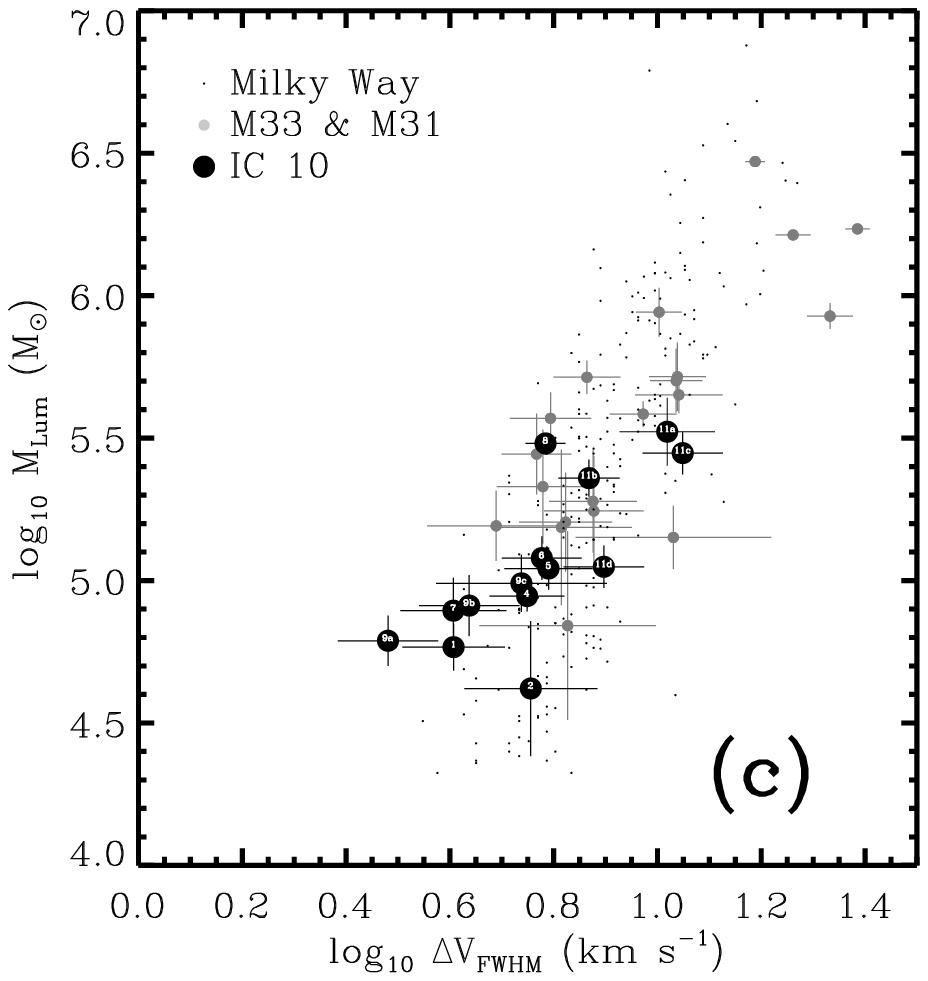}

\figcaption{\label{LARSON} Larson's laws for clouds in IC~10 (black
circles), compared to the Milky Way (black dots) and the Local Group
spirals M~31 and M~33 (gray circles). (a) The size-velocity dispersion
relation.  (b) The virial mass-luminosity relationship. (c) The
luminosity-line width relationship. All four galaxies display very
similar behavior, obeying the same scaling of their global GMC
properties.}

\end{figure*}

\begin{figure*}
\epsscale{0.45} 
\plotone{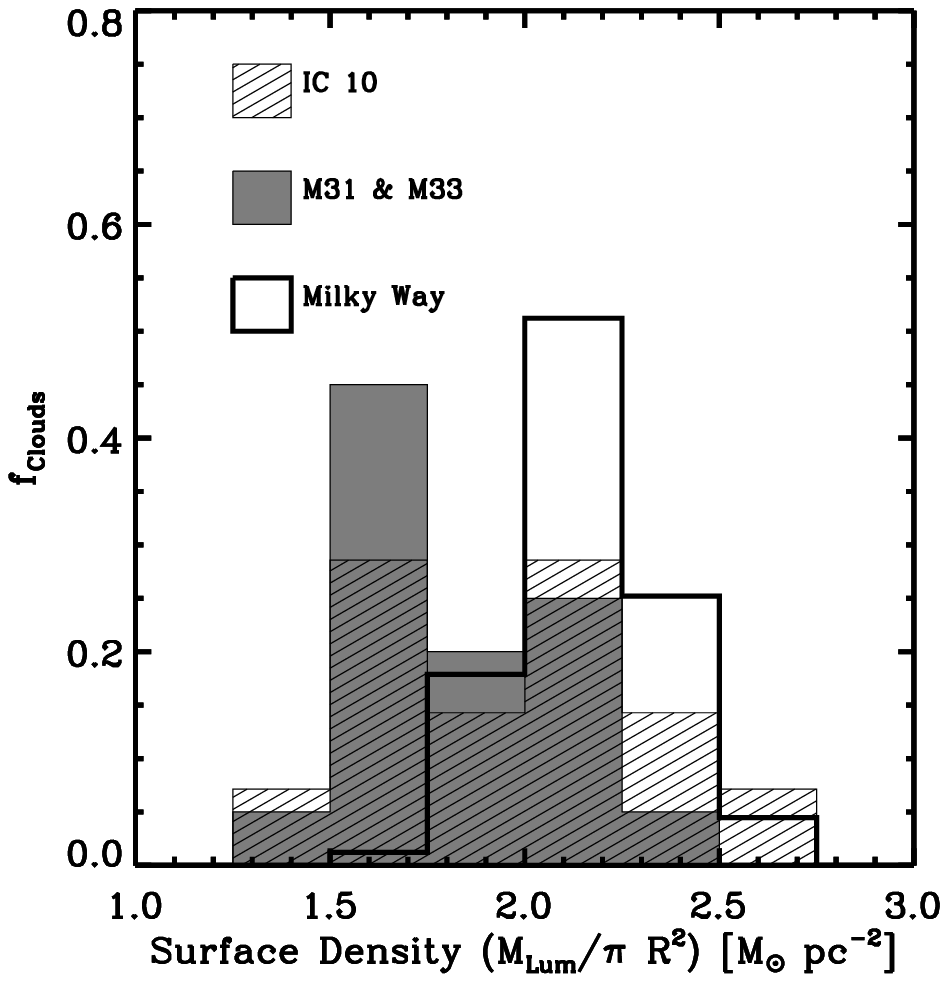} \plotone{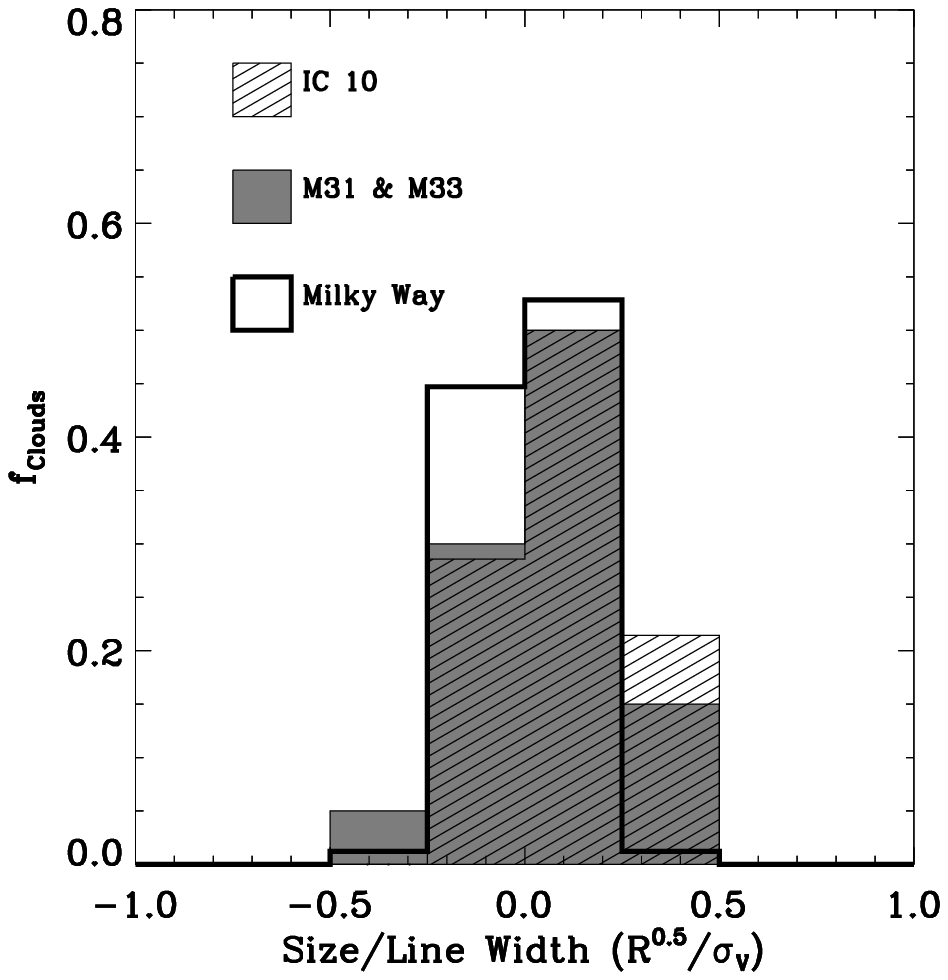}
\plotone{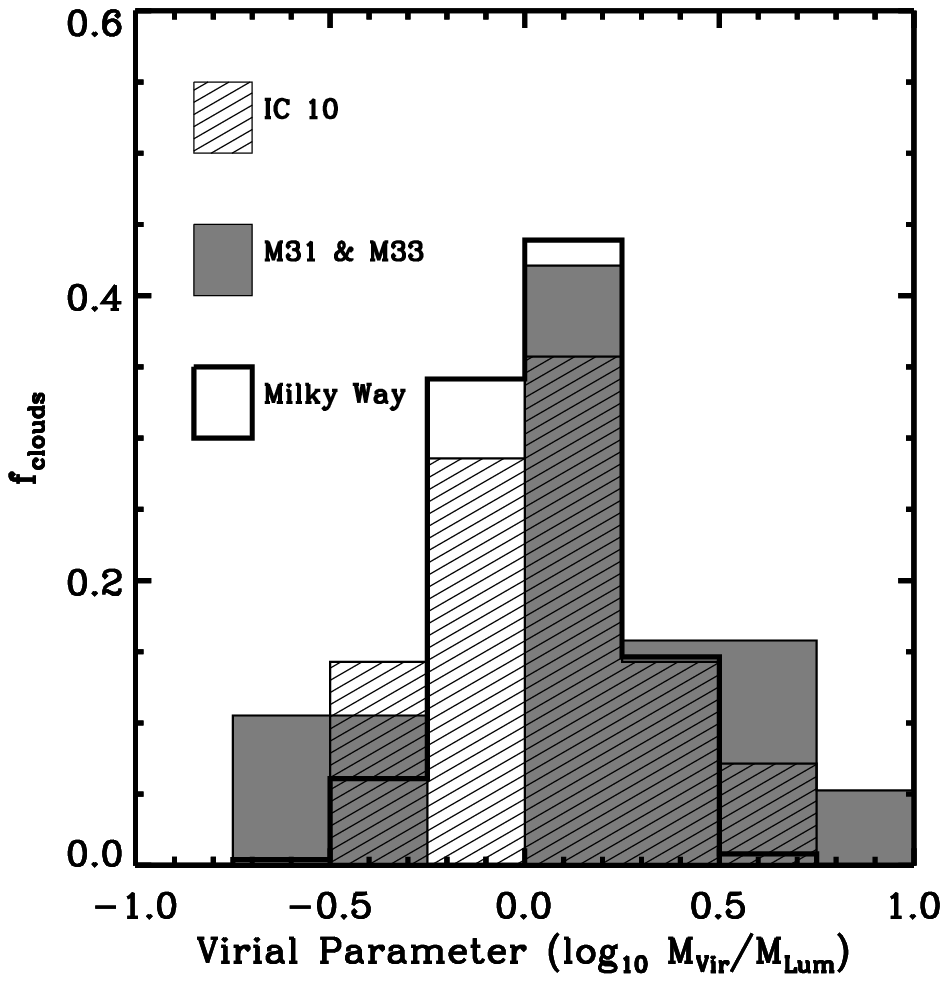} 
\figcaption{\label{LARSONHIST} Distributions around the Larson
relationships for clouds in IC~10 (hashed), the Milky Way (thick
line), and M~31 and M~33 (gray histogram), showing the similarities
among the GMC populations in the four galaxies.}
\end{figure*}

Figure \ref{LARSON} shows that clouds in IC~10 lie on or near the
scaling laws for GMCs in the Milky Way, M~31, and M~33. This
conclusion is reinforced by Figure \ref{LARSONHIST}, which shows three
physical parameters --- the surface density, the virial to luminous
mass ratio, and the scatter about the size-velocity dispersion
relation ($\sigma_v \propto R^{0.5}$) --- for clouds in IC~10, the
Milky Way, M~31, and M~33. Table \ref{KSTESTS} show the results of two
sided KS-tests comparing these physical parameters between IC~10 and
the Milky Way (column 2) and M~31+M~33 (column 3). The distribution of
physical parameters for IC~10 GMCs agrees very well with the
distribution found for the M~31 and M~33 GMCs ($<2\sigma$ difference
for each parameter). Since the M31/M33 GMC properties are also derived
from interferometric data and are the result of the same analysis used
to produce the IC~10 GMC properties, we attach particular weight to
this comparison. The KS test detects $\sim 3\sigma$ differences
between in IC~10 the Milky Way in the both surface density of clouds
and their scatter about the size-velocity dispersion relation. Figure
\ref{LARSONHIST} shows that these differences are relatively small
however and since the same differences exist between the Milky Way
data and the M~31/M~33 data they may represent a systematic difference
stemming from either observational biases or differences in the
measurement algorithm. {\em The CO-emitting clouds resolved by OVRO in
IC~10 are very similar to GMCs in the Milky Way. They are
indistinguishable from GMCs in M~31 and M~33 observed and analyzed in
the same manner as the IC~10 clouds.}

\begin{deluxetable*}{l c c}
\tabletypesize{\small}
\tablewidth{0pt}
\tablecolumns{5}
\tablecaption{\label{KSTESTS} KS Test Comparisons of IC~10 GMCs}

\tablehead{ \colhead{Property} & \colhead{vs. Milky Way} & \colhead{vs. M~31+M~33}}

\startdata
Surface Density & $0.02$\tablenotemark{a} & $0.65$ \\
Size $^{0.5}$ / Velocity Dispersion & $1 \times 10^{-3}$ & $0.41$ \\
Virial Parameter & $0.25$ & $0.67$ \\
\enddata 
\tablenotetext{a}{Entries are the probability of generating
the observed difference or greater from the same parent population
(i.e. randomly).}
\end{deluxetable*}

\subsection{The Mass Spectrum of GMCs in IC~10}

\begin{figure}
\epsscale{1.0} 
\plotone{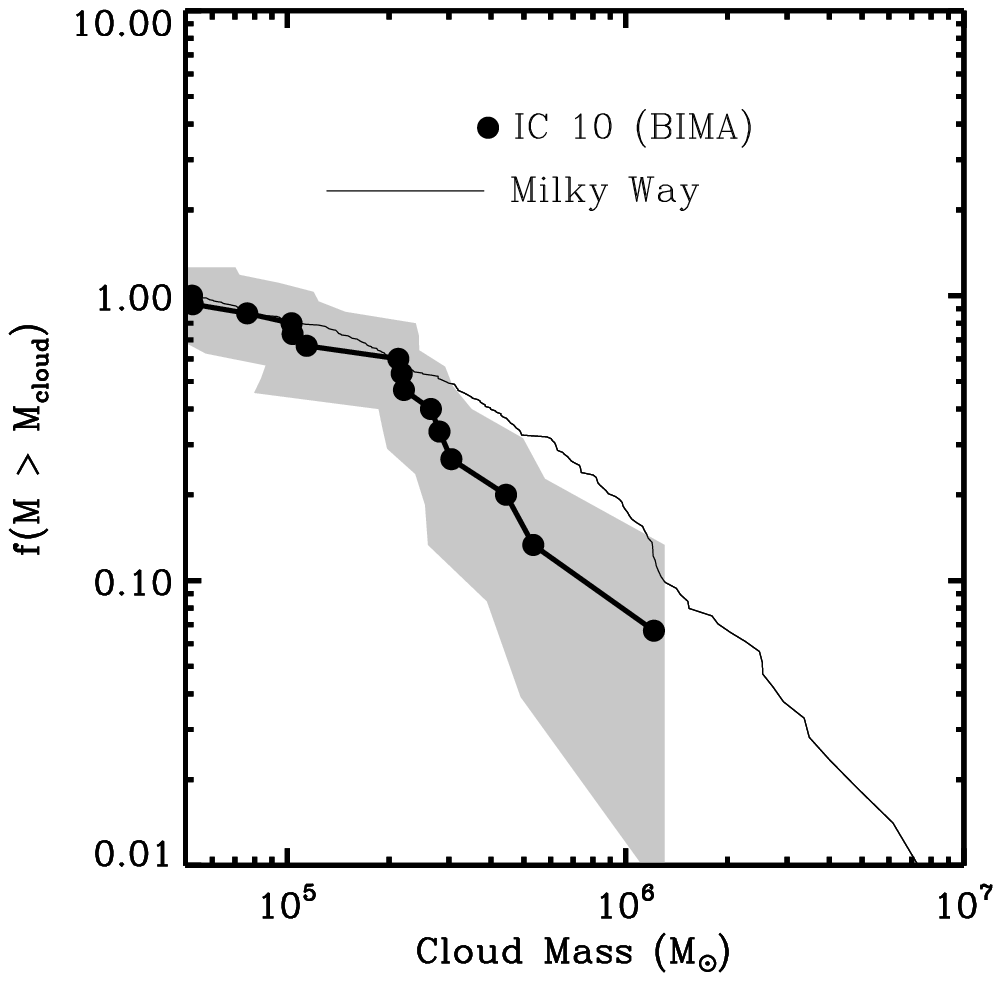}

\figcaption{\label{CDF} The cumulative distribution function for GMCs
in the Milky Way (solid line) and IC~10 (black circles). We include
only Milky Way clouds with $M > 10^5 M_{\odot}$ (the approximate
sensitivity of our survey). The gray region shows the area of
$1\sigma$ uncertainty, but systematic effects from the identification
of clouds are also more important and are not shown.}
\end{figure}

Do GMCs in IC~10 also exhibit the same {\em distribution} of masses as
GMCs in other galaxies? We calculate the cumulative distribution
function (CDF) for GMC masses in our data (calculated from the GMC
luminosities), so that the value of the CDF at a particular mass is
equal to the fraction of clouds with masses greater than or equal to
that mass. The power law index of the CDF indicates how ``top heavy''
or ``bottom heavy'' the distribution of cloud masses is. Figure
\ref{CDF} shows the CDF for clouds from the BIMA survey and GMCs above
$5 \times 10^4$ M$_{\odot}$ from \citet[][]{SOLOMON87}. The best fit
power law to the BIMA survey CDF yields $\frac{dN}{dM} \propto M^{-2.0
\pm 0.2}$, a spectrum with approximately equal mass in each
logarithmic bin. This index agrees with the Milky Way slope we find
when we consider the \citet[][]{SOLOMON87} data over the same range of
masses ($-1.85 \pm 0.05$), but is steeper (more ``bottom heavy'') than
the slope of $\approx -1.6$ derived from all of the
\citet[][]{SOLOMON87} data. The OVRO data also yield a power law index
of $\sim -2$ (though they do not represent a complete sample and show
a normalization offset). The exact power law index of the mass
distribution in IC~10 is quite uncertain because it depends heavily on
the identification of GMCs in the BIMA survey (for example the large
cloud in the southeast is resolved into three separate, smaller
clouds, by OVRO). By altering our decomposition of the survey, we are
able to obtain power law indices between $-1.8 \pm 0.2$ and $-2.2 \pm
0.2$. The gray region in Figure \ref{CDF} shows uncertainties
associated with counting errors and uncertainties in the mass, but not
decomposition.

We have tested for consistency by realizing $1000$ test samples of
Milky Way GMCs and comparing their mass distribution to that of the
IC~10 GMCs. Each test sample contains $16$ clouds (to match the IC~10
sample) randomly drawn from the population of \citet[][]{SOLOMON87}
clouds with masses $> 5 \times 10^4~$M$_{\odot}$, allowing repeats. We
compare each test sample to the population of clouds in IC~10 using a
two sided KS test. As a control, we perform the same test using pairs
of test samples from the Milky Way data. The median comparison of
IC~10 GMCs to Milky Way GMCs (over the $1000$ test samples) showed
more difference than $65\%$ of the control comparisons. Any
differences between the two populations are thus of only $\sim
1\sigma$ significance.

A mass distribution with a power law index of $-2$ implies a
significant amount of gas below our completeness limit. As mentioned
in the previous section, these small clouds might explain the
discrepancy between the single dish and interferometer results if they
exist in an extended layer (say, throughout the \hi\ filaments) near
the large GMCs.

\subsection{CO and \ion{H}{1}}

\begin{figure}
\epsscale{1.0} 
\plotone{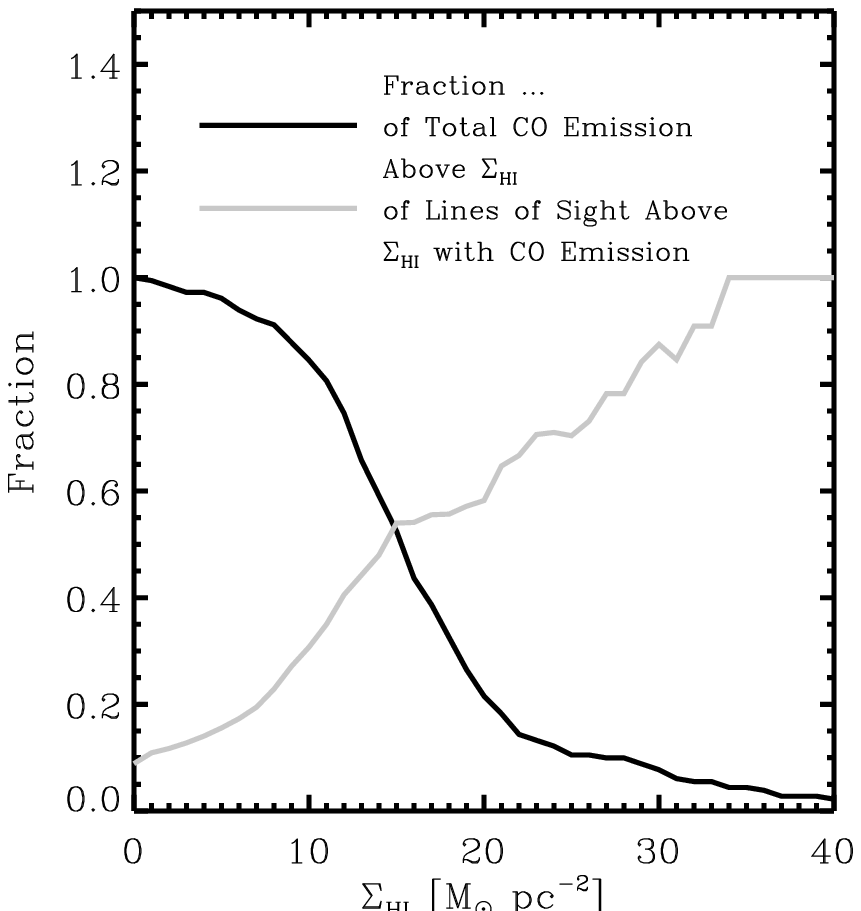}

\figcaption{\label{COHIFRAC} The fraction of CO emission from IC~10
above a given HI column density, $\Sigma_{HI}$ is shown in black. The
fraction of lines of sight with a given HI column density that also
show CO emission is shown in gray. Most of the emission comes from
regions of the galaxy with column densities above 10 M$_{\odot}$
pc$^{-2}$, and $50\%$ of the emission arises from regions with
$\Sigma_{HI}\gtrsim16$~\sdunits. Only $50\%$ of lines of sight with
this column density has associated molecular gas, however. Although
$\Sigma_{HI} \gtrsim 10$ is a necessary condition to find molecular
gas, it is not sufficient on its own.}

\end{figure}

The GMCs in IC~10 are found almost exclusively in high column density
atomic gas filaments.  Figure \ref{COMAP} shows that we find molecular
gas only where we find atomic gas, usually above a surface density of
$\Sigma_{HI} = 10$ M$_{\odot}$ pc$^{-2}$ ($N(\hi) = 1.25 \times
10^{21}$ cm$^{-2}$). Figure \ref{COHIFRAC} shows the relationship
quantitatively, plotting the fraction of the total CO emission as a
function \hi\ column density along the line of sight (in black) and
the fraction of lines of sight with the specified column density that
have associated CO emission (in gray). Half of the molecular gas
emission comes from regions with \hi\ column densities above $15$
M$_{\odot}$ pc$^{-2}$ and $85 \%$ is found above a contour of
$\Sigma_{HI} = 10$ M$_{\odot}$ pc$^{-2}$. A high column of atomic gas
is not a sufficient condition, though, as only $30\%$ of the area in
IC~10 with $\Sigma_{HI} > 10$ M$_{\odot}$~pc$^{-2}$ has associated
molecular gas and only half of the lines of sight with $\Sigma_{HI} =
16$ M$_{\odot}$ pc$^{-2}$ ($N(\hi) = 2.0 \times 10^{21}$ cm$^{-2}$)
have associated CO emission. This discrepancy (necessary, but not
sufficient) arises mostly as a result of the large region of
relatively high column atomic gas to the east of IC~10. This high
column gas is mostly devoid of CO emission, harboring only one
molecular cloud. \citet[][]{ENGARGIOLA03} found a similar result in
their survey of M~33 --- the CO emission comes almost exclusively from
within the \hi\ filaments but the presence of a filament does not
necessarily imply the presence of CO emission. Similar results can be
seen comparing CO emission from the LMC to \hi\
\citep[][]{FUKUI99,KIM98}. Broadly, this is the same effect seem in
many disk galaxies: the \hi\ extends far out into the disk while
molecular gas and star formation are relatively centrally confined
\citep[e.g.][]{WONG02}.

\citet[][]{WONG02} and \citet{BLITZ04} have suggested that the
hydrostatic gas pressure, $P_h$, can predict the ratio of molecular to
atomic gas, $f_{mol} = \Sigma_{mol} / \Sigma_{HI}$, over a region of a
galaxy. The hydrostatic gas pressure may trace the volume density of
gas, $\rho_{gas}$, since $P_h = \rho_{gas}~v_g^2$ and the gas velocity
dispersion, $v_g$, is often quite constant \citep[][and references
therein]{BLITZ04}. The volume density of gas should be more relevant
to the formation of H$_2$ from \hi\ than the surface density. We test
whether these arguments hold in IC~10 by calculating $P_h$ using the
formula derived by \citet[][]{BLITZ04} for a stellar-dominated disk,

\begin{equation}
P_{h} =  0.84 (G \Sigma_*)^{0.5}\Sigma_g \frac {v_g} {(h_*)^{0.5}} \
\label{approxpressure}
\end{equation}

\noindent where $\Sigma_{g}$ is the total surface density of the gas,
$\Sigma_{*}$ is the surface density of stars, $v_g$ is the velocity
dispersion of the gas, and $h_*$ is the scale height of the stellar
disk. From the \hi\ cube \citep[][]{WILCOTS98}, we measured the median
$v_g$ across the disk (for lines of sight with non-negligible \hi\
emission) to be $\approx 7.5$ km s$^{-1}$. We assume the scale height
of stars to be comparable to that of disk galaxies \citep[$\sim 300$
pc, see][]{BLITZ04} and calculate $f_{mol}$ using $X_{CO} = 2 \times
10^{20}$ \xcounits\ (appropriate to the GMCs but perhaps not diffuse
gas, see \S\ref{XCOSECT}).

\begin{figure}
\epsscale{1.0} 
\plotone{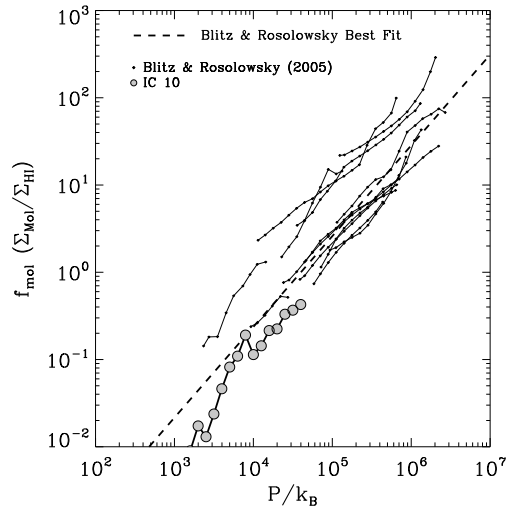}

\figcaption{\label{PRESS} The molecular-to-atomic ratio plotted as a
function of midplane pressure in IC~10. Results and the best fit
relation for spiral galaxies from Blitz \& Rosolowsky (2005) are shown
for comparison. We bin data with similar pressures along lines of
sight with $\Sigma_* > 50$ M$_{\odot}$ pc$^{-2}$, where the disk
approximation may be valid, and apply an inclination correction of
$0.67$.}

\end{figure}

Figure \ref{PRESS} shows reasonable agreement between the IC~10 data
and an extrapolation from the \citet[][]{BLITZ05} galaxies to low
pressure. $P_h$ and $f_{mol}$ have a rank correlation coefficient of
$0.7 \pm 0.1$. The median ratio, $P_0 = \frac{P_h}{f_{mol}}$, is $7
\times 10^4~k_B$ and is uncertain by a factor of two. This value of
$P_0$ (the pressure for which $f_{mol} = 1$) is consistent with the
results of \citet[][]{BLITZ05}, who find $P_0 = 4 \times 10^4~k_B$ and
a nearly linear best fit relation. On average IC~10 is poorer in
molecular gas than one might expect by about a factor of two \citep[it
lies just under the best fit line from][]{BLITZ05}. Since we used the
BIMA survey in this comparison, the discrepancy might be almost
completely negated by including the resolved out flux.

Table \ref{RANKCORR} shows that the pressure is a better predictor of
$f_{mol}$ than either the atomic gas surface density or the stellar
surface density. The rank correlation between $P_h$ and $f_{mol}$ is
$0.7 \pm 0.1$, higher than the rank correlation between $f_{mol}$ and
either $\Sigma_{HI}$ or $\Sigma_{*}$. Both $\Sigma_{HI}$ and $P_h$ are
very highly correlated with $\Sigma_{mol}$, which demonstrates that
the conclusion of \citet[][]{BLITZ05} seems to hold in IC~10: \hi\ is
a necessary prerequisite for the presence of molecular gas, but not
the best predictor of the ratio of molecular to atomic gas. The
hydrostatic gas pressure offers an improved prediction of the ratio of
molecular to atomic gas because the gravitational influence of a
stellar potential well is necessary to enable the formation of H$_2$
out of \hi\ filaments.

\begin{deluxetable*}{l c c}
\tabletypesize{\small}
\tablewidth{0pt}
\tablecolumns{5}
\tablecaption{\label{RANKCORR} Rank Correlations with Molecular to
  Gas Content}

\tablehead{ \colhead{Property} & \colhead{Rank Correlation with
    $\Sigma_{mol}$} & \colhead{Rank Correlation with
    $\Sigma_{mol}$/$\Sigma_{HI}$}}

\startdata
$\Sigma_{HI}$ & $0.6 \pm 0.1$ & $0.4 \pm 0.1$ \\
$\Sigma_{*}$ & $0.4 \pm 0.1$ & $0.5 \pm 0.1$ \\
$P_h$ & $0.8 \pm 0.1$ & $0.7 \pm 0.1$ \\
\enddata
\end{deluxetable*}

\subsection{Star Formation and Gas in IC~10}
\label{SFRSEC}

We showed above that GMCs in IC~10 are similar to those in the Milky
Way, M~31, and M~33. Does the molecular gas also form stars at the
same rate as molecular gas in spirals? In this section we compare the
star formation efficiency (SFR per unit gas) in IC~10 to that in
larger galaxies. We perform these comparisons using the \hi, CO, and
H$\alpha$ maps convolved to a common spatial resolution of $275$ pc
($60''$, with $30''$ pixels). This is comparable to the scale height
of the stellar disk in a spiral galaxy, so we expect the averaging to
occur over roughly the same distance along the line of sight and
perpendicular to the line of sight. We limit ourselves to regions of
the stellar disk of IC~10 with stellar surface densities in excess of
$\Sigma_*\sim50$ M$_\odot$~pc$^{-2}$ (obtained from the $K$-band
light), because outside this region the three dimensional structure of
the \hi\ envelope is uncertain and it is unclear how to interpret the
line of sight surface densities.

\begin{figure}
\epsscale{1.0} 
\plotone{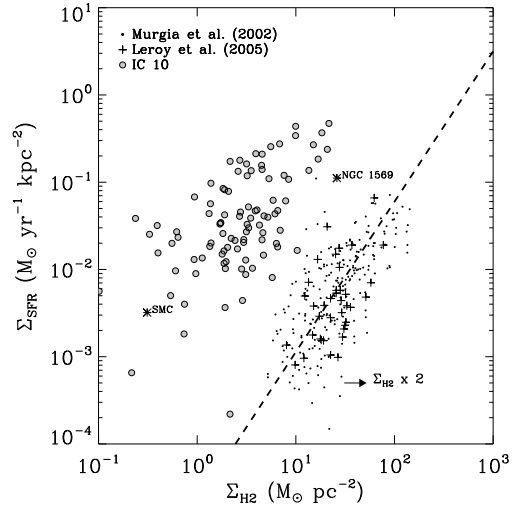}

\figcaption{\label{COSFR} Star formation surface density as a function
of H$_2$ surface density in IC~10. Global values for large galaxies
\citep[][]{MURGIA02} and LMC-size dwarfs \citep[][]{LEROY05} are shown
for comparison. We convolve the data to a resolution of $1'$ (275 pc)
and apply an inclination correction of $0.67$. We plot only points in
the stellar disk of IC~10, where $\Sigma_* > 50$ M$_{\odot}$
pc$^{-2}$. IC~10 has much more star formation per unit molecular gas
than most spirals or dwarfs. We highlight two other dwarfs similar to
IC~10: NGC~1569 and the SMC.}

\end{figure}

For galaxies the size of the LMC or bigger, the efficiency with which
stars form from molecular gas (and its inverse, the molecular gas
depletion time) depends wealky on galaxy mass and Hubble type
\citep{YOUNG91,YOUNG95,MURGIA02,LEROY05}. Figure \ref{COSFR} shows the
relationship between surface density of star formation and molecular
gas surface density for a range of galaxies. Both dwarf galaxies and
spirals obey a power law relationship of roughly

\begin{equation}
\Sigma_{SFR} = 10^{-3.4 \pm 0.1} \Sigma_{Mol}^{1.3 \pm 0.1} \mbox{ ,}
\end{equation}

\noindent where $\Sigma_{SFR}$ is the star formation surface density
in units of M$_{\odot}$ yr$^{-1}$ kpc$^{-2}$ and $\Sigma_{Mol}$ is the
molecular gas surface density in units of M$_{\odot}$
pc$^{-2}$\citep[][]{MURGIA02,LEROY05}. Figure \ref{COSFR} also shows
that IC~10 very clearly does {\em not} fall on this trend. Rather, the
data for IC~10 show much larger rates of star formation per unit
molecular gas than is found in other galaxies --- a median factor of
$270$ higher. IC~10 is not unique in this regard: both the SMC
\citep[][]{MIZUNO01,WILKE04} and the nearby starburst NGC~1569 also
show higher SFR surface densities than their CO content would suggest
unless \xco\ is considerably larger than Galactic in these sources.

\begin{figure}
\epsscale{1.0} 
\plotone{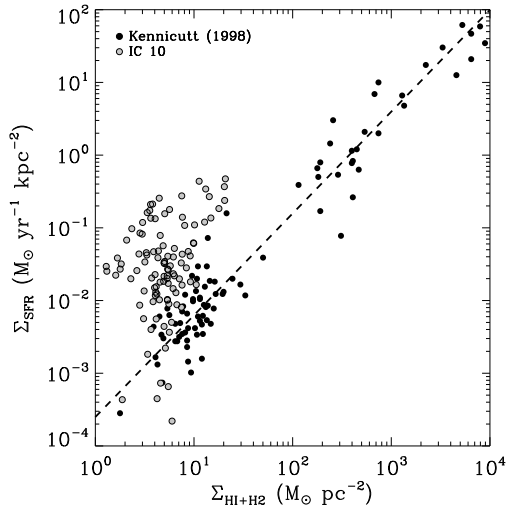}

\figcaption{\label{SLAW} The Schmidt Law in IC~10. The points and
relation of \citet[][]{KENNICUTT98} are shown for comparison. The data
have been convolved to a resolution of $1'$ (275 pc) and an
inclination correction of $0.67$ is applied. We plot only points in
the stellar disk of IC~10, where $\Sigma_* > 50$ M$_{\odot}$
pc$^{-2}$. IC~10 shows more star formation per unit gas than large
galaxies.}

\end{figure}

The total gas surface density, rather than the H$_2$ surface density,
is often used as a predictor of the star
formation. \citet[][]{KENNICUTT98} found that a single power law
described the relationship total gas surface density and the star
formation surface density across a wide range of galaxies. Data from
that paper are plotted along with gas surface densities and H$\alpha$
derived star formation surface densities (SFSDs) in Figure
\ref{SLAW}. Figure \ref{SLAW} shows that in IC~10, the highest gas
surface densities do roughly correspond to the highest star formation
rates, although the agreement with \citet[][]{KENNICUTT98} is poor. In
particular, regions of high SFSD seem to have lower gas surface
densities in IC~10 than galaxies with the same SFSD from the
\citet[][]{KENNICUTT98} sample. Furthermore, the scatter in SFSD for a
given total gas surface density is very large, particularly around
$\Sigma_{HI+H2} \approx 7$ M$_{\odot}$ pc$^{-2}$.

How can we explain the extremely high efficiency of star formation in
IC~10 given the similarity between its GMC properties and those of
spiral GMCs? We suggest several possibilities.

\noindent 1. IC~10 has much more molecular gas than we infer from the
CO. The discrepancy between IC~10 and large galaxies may be entirely
explained away by a factor of $\sim 30$ increase in the CO-to-H$_2$
conversion factor. We have presented evidence above that BIMA may
resolve out as much flux as it recovers (leaving us with only a factor
of $\sim 15$ discrepancy), but the GMC properties we measured above
suggest values of $X_{CO}$ that are nearly Galactic. Adjusting the
data in Figures \ref{COSFR} and \ref{SLAW} to agree with other
galaxies requires more than just a large reservoir of hidden molecular
gas; despite producing relatively small amounts of CO emission, this
molecular gas must be associated with star formation.  It is unlikely
that the physical conditions in a hidden reservoir of H$_2$ could
simultaneously be conducive to star formation, inhospitable to CO, and
not contribute to the virial masses of the GMCs.

\noindent 2. The star formation rate of IC~10 is lower than we
estimate from the H$\alpha$ due to line-blanketing effects in stellar
atmospheres. These effects would cause us to underpredict the UV
radiation generated per star, leading to an overestimate of the star
formation rate.  Studying a sample of stars in the SMC,
\citet[][]{MASSEY05} found that O stars in that system (with $Z \sim
0.1 Z_{\odot}$) had effective temperatures $\sim 4000$ K higher than
their solar metallicity analogues. For an O5 star ($T_{eff} \sim
45,000$), this results in the SMC star producing $\sim 50\%$ more
ionizing photons. A similar result is obtained using population
synthesis codes such as STARBURST99 \citep[Figure 78 in
][]{STARBURST99} --- a shift of an order of magnitude in metallicity
for a continuously star forming system results in an increase of $\sim
50\%$ in the number of ionizing photons produced. These adjustments
are not large enough to make a significant dent in the discrepancy
between IC~10 and larger galaxies.

\noindent 3. If IC~10 has an unexpectedly top heavy IMF our inferred
star formation rates may be too high. \citet[][]{KENNICUTT94} explores
the effect of adopting other (Milky Way) IMFs (such as those of
\citet[][]{SCALO86} or \citet[][]{KTG93}) and finds increases of a
factor of $\sim 2-3$ to the star formation rate per unit H$\alpha$
emission. A very top-heavy IMF, by contrast, would have the effect of
lowering the amount of star formation per unit H$\alpha$
luminosity. The lion's share of ionizing photons are produced by stars
with $M > 10$ M$_{\odot}$. For the Salpeter IMF assumed in the
H$\alpha$ calculation \citep[][]{KENNICUTT94}, $\sim 10\%$ of the mass
of stars resides in stars with $M > 10$ M$_{\odot}$. If IC~10 produced
{\em only} these stars then the star formation efficiency might be a
factor of 10 lower than the value we calculate here. A very top heavy
IMF could also explain the unusually high abundance of Wolf Rayet
stars in IC~10. However, an IMF that is dramatically skewed towards
high masses would contradict the finding by \citet[][]{HUNTER01} that
the clusters in IC~10 are consistent with a Galactic IMF. Barring such
a dramatic IMF, it seems unlikely that the offset we observe is only a
result of a miscalibration of H$\alpha$ as a star formation tracer.

\noindent 4. The star formation rate in IC~10 may have been higher in
the recent past. If IC~10 has recently undergone a period of intense
star formation and is now forming stars at (relatively) more modest
rate, we may be catching it at a point its life-cycle during which it
is (relatively) depleted in molecular gas but still showing the signs
of a recent star burst. This scenario could also explain the very high
WR star counts and perhaps the discrepancy between the various star
formation tracers. In this case, IC~10's high star formation
efficiency may be temporary, an artifact of {\em when} we are
observing the galaxy. In such a case we would expect a large sample of
IC~10-like dwarfs to average to a position consistent with the other
galaxies in figures \ref{COSFR} and \ref{SLAW}. The higher star
formation rate must have occurred within the last $\sim 10$~Myr
because high mass stars formed during the period of higher SFR must
still be contributing UV photons that create the H$\alpha$ flux. This
is consistent with the ages of the clusters found by
\citet[][]{HUNTER01}, 4 -- 30 Myr.

\noindent 5. Finally, the efficiency of star formation within
molecular clouds may indeed be higher in IC~10 than in the Milky Way
or other galaxies. This result seems to contradict the similarities
between GMCs in IC10 and those in M~31, M~33, and the Milky
Way. However, we have emphasized the environmental differences between
IC~10 and these systems and these differences may be manifesting
themselves in an unexpected way that dramatically affects the
efficiency with which molecular gas forms stars.

% CONCLUSIONS AND SPECULATION
\section{Summary and Conclusions}

We present a complete survey $^{12}$CO $J=$\jone\ in IC~10 using
BIMA. The survey covers all of the optical disk of IC~10 and a large
part of the extended \hi\ structure with a resolution of $14''$ (70
pc) and a sensitivity sufficient to detect clouds with masses greater
than $4 \times 10^4$ M$_{\odot}$. We find structures across the
optical disk of the galaxy and in the extended structure to the north
and east of the galaxy. The BIMA finds a total CO luminosity of $1.0
\times 10^6$ K km s$^{-1}$ pc$^2$.

We also present ARO 12m observations of 22 fields in IC~10,
corresponding to most of the locations in which \co\ emission is
detected by the BIMA survey and a number of locations of interest with
no \co\ emission. The ARO 12m detects \co\ emission only where BIMA
also detects \co\ emission. The ARO 12m finds more emission than BIMA
or OVRO along the same line of sight. This may be evidence for an
extended \co\ component surrounding the more compact structures
detected by the interferometers. Comparing the integrated luminosity
from the ARO 12m to the BIMA survey, we estimate that the true CO
luminosity of IC~10 is $2.2 \times 10^6$ K km s$^{-1}$ pc$^2$.

We measure the properties of 14 resolved CO structures in IC~10 from
high resolution OVRO data by \citet[][]{WALTER03}. The sizes, line
widths, and luminosities of these structures resemble those of GMCs
found in similar surveys of M~31, M~33 \citep[][]{ROSOLOWSKY03,
ROSOLOWSKY05}, and Milky Way GMCs \citep[][]{SOLOMON87}. We conclude
that we are observing GMCs in IC~10. The virial-mass to luminosity
ratio in these GMCs is comparable to that observed for spiral galaxy
GMCs and we argue that this implies that a Galactic CO-to-H$_2$
conversion factor applies to IC~10. We can not constrain the
conversion factor in the gas resolved out by the interferometers.

Most of the CO emission detected by the BIMA survey comes from lines
of sight with $\Sigma_{HI}$ above 10 M$_{\odot}$ pc$^{-2}$ ($N(\hi) =
1.25 \times 10^{21}$ cm$^{-2}$) and all of the CO emission is very
close to such regions. However, only 30$\%$ of the lines of sight with
$\Sigma_{HI} > 10$ M$_{\odot}$ pc$^{-2}$ have associated CO
emission. This may be because not all high column density \hi\
actually corresponds to high volume density gas. Indeed, we show that
hydrostatic gas pressure predicts the CO along a line of sight better
than the \hi\ column alone. Further, the ratio of molecular to atomic
gas along a line of sight obeys the same simple relationship in IC~10
that is seen in large spiral galaxies.

IC~10 displays significantly more star formation (H$\alpha$) for a
given gas surface density (molecular or total) than large spiral
galaxies. We suggest several explanations for this: a higher
CO-to-H$_2$ conversion factor, a different IMF or SFR calibration, or
a biased (in time) perspective. The last explanation is appealing ---
namely that IC~10's unusual star formation efficiency is a timing
effect. We may be observing a starburst just past its peak, so that
the signatures of star formation are still present but the
star-forming gas has already been somewhat depleted.

\acknowledgements This research was supported by NSF grant
AST-0228963. We thank Eric Wilcots and Brian Miller for sharing their
VLA \hi\ cubes; Steve Dawson for contributing his Hat Creek summer
school time to fill in the D array survey; Erik Rosolowsky for
valuable comments on a draft of this paper; Armando Gil de Paz and Tom
Jarrett for sharing optical and near-IR data; Josh Simon for help with
the $K$-band data; Tam Helfer and Tony Wong for code and help with
data reduction; Jason Wright and Ryan Chornock for help with input on
signal detection and data presentation; Lucy Ziurys the staff of the
ARO 12m, in particular Paul Hart and Aldo Apponi, for assistance in
taking the ARO 12m data. This paper makes extensive use of: the
NASA/IPAC Extragalactic Database (NED) which is operated by the Jet
Propulsion Laboratory, California Institute of Technology, under
contract with the National Aeronautics and Space Administration;
NASA's Astrophysics Data System (ADS); and data products from the Two
Micron All Sky Survey, which is a joint project of the University of
Massachusetts and the Infrared Processing and Analysis
Center/California Institute of Technology, funded by the National
Aeronautics and Space Administration and the National Science
Foundation.

\end{document}